\documentclass[
  journal=pasa,
  manuscript=research-paper, %% or "review"
  year=2024,
]{cup-journal}

\usepackage{microtype,siunitx,booktabs}
\usepackage{hyperref} 
\hypersetup{colorlinks,citecolor=blue,linkcolor=blue,urlcolor=blue}

\newcommand{\maUpNum}{10000}

\newcommand{\maGnuastroVersion}{0.15}
\newcommand{\maNoiseChiselkernel}{-{}-kernel=none}
\newcommand{\maNoiseChiselqthresh}{-{}-qthresh=0.5}
\newcommand{\maNoiseChiselsnquant}{-{}-snquant=0.999}
\newcommand{\maNoiseChiseldetgrowquant}{-{}-detgrowquant=0.85}

\newcommand{\maDiffuseAllSig}{9.7}

\newcommand{\maCatHiResFiltozzN}{0.20}
\newcommand{\maCatHiResFiltozzNe}{0.08}
\newcommand{\maCatHiResFiltozzB}{0.11}
\newcommand{\maCatHiResFiltozzBe}{0.07}
\newcommand{\maCatHiResFiltozzS}{0.11}
\newcommand{\maCatHiResFiltozzSe}{0.05}

\newcommand{\maCatHiResDifeee}{4.88}
\newcommand{\maCatHiResDifeeee}{0.50}
\newcommand{\maDiffuseSigmaOne}{6.6}
\newcommand{\maDiffuseSigmaTwo}{7.1}

\newcommand{\maCatHiResFileeeNe}{0.11}

\newcommand{\maCatHiResFileeeBe}{0.10}

\newcommand{\maCatHiResFileeeSe}{0.07}

\newcommand{\maCatLowReseee}{18.9}

\newcommand{\maCatLowResSpecIndP}{0.22}
\newcommand{\maCatLowResSpecIndN}{0.19}

\usepackage{graphicx}
\usepackage{amsmath}
\usepackage{amssymb}
\usepackage{booktabs}
\usepackage{epstopdf}
\usepackage{float}
\usepackage{soul}
\sethlcolor{yellow}
\usepackage{xcolor}
\usepackage{ulem}
\usepackage{color}
\usepackage[nolist]{acronym}
\usepackage{epstopdf}
\usepackage{subcaption}
\usepackage{siunitx}
\usepackage{gensymb}
\usepackage{footnote}
\usepackage{comment}
\usepackage{tablefootnote}
\usepackage[]{threeparttable}
\usepackage{tabto}    
\usepackage{pdfpages}
\usepackage[switch,mathlines,pagewise]{lineno}

\newcommand{\lsun}{\ifmmode{{\rm ~L}_\odot}\else{~L$_\odot$}\fi}
\newcommand{\Msun}{\ifmmode{{\rm ~M}_\odot}\else{~M$_\odot$}\fi}

\newcommand{\ujybm}{$\mu$Jy~beam$^{-1}$\,}

\newcommand{\degr}{$^{\circ} $}
\newcommand{\arcmin}{$^{\prime}$}
\newcommand{\arcsec}{$^{\prime\prime}$}

\graphicspath{./images/} %Setting the graphicspath

\begin{acronym}
    \acro{AGN}      {Active Galactic Nuclei}
    \acro{APEC}     {Astrophysical Plasma Emission Code}
    \acro{ASKAP}    {Australian Square Kilometre Array Pathfinder}
    \acro{ATCA}     {Australia Telescope Compact Array}
    \acro{ATNF}     {Australia Telescope National Facility}
    \acro{BCG}      {Brightest Cluster Galaxy}
    \acro{carta}    {Cube Analysis and Rendering Tool for Astronomy}
    \acro{CXO}      {Chandra X-ray Observatory}
    \acro{DSS}      {Digitised Sky Survey}
    \acro{eeHIFLUGCS} {extremely expanded HIghest X-ray FLUx Galaxy Cluster Sample}
    \acro{EMU}      {Evolutionary Map of the Universe}
    \acro{FOV}      {Field of View}
    \acro{FWHM}     {Full Width at Half-Maximum}
    \acro{FRI}      {Fanaroff-Riley Type I}
    \acro{GMRT}     {Giant Metrewave Radio Telescope}
    \acro{ICM}      {Intra-Cluster Medium}
    \acro{LOFAR}    {Low Frequency Array}
    \acro{MCXC}     {Meta-Catalogue of X-ray Detected Clusters of Galaxies}
    \acro{MRO}      {Murchison Radio-astronomy Observatory}
    \acro{MWA}      {Murchison Widefield Array}
    \acro{NRAO}     {National Radio Astronomy Observatory Very Large Array Sky Survey}
    \acro{NVSS}     {NRAO VLA Sky Survey}
    \acro{PAF}      {Phased Array Feed}
    \acro{PAWSEY}   {Pawsey Supercomputing Centre}
    \acro{RACS}     {Rapid \ac{ASKAP} Continuum Survey}
    \acro{RFI}      {Radio Frequency Interference}
    \acro{SED}      {Spectral Energy Distribution}
    \acro{SUMSS}    {Sydney University Molonglo Sky Survey}
    \acro{TGSS}     {\ac{TIFR} \ac{GMRT} Sky Survey}
    \acro{TIFR}     {Tata Institute of Fundamental Research}
    \acro{uGMRT}    {upgraded Giant Metrewave Radio Telescope}
    \acro{WAT}      {Wide-Angle-Tail}
    \acro{WISE}     {Wide-field Infrared Survey Explorer}
    \acro{XMM}      {XMM-Newton}
\end{acronym}

\title[EMU Observations of Abell~S1136]{Evolutionary Map of the Universe (EMU): Observations of Filamentary Structures in the Abell~S1136 Galaxy Cluster}

\author{Peter. J. Macgregor}
\affiliation{School of Science, Western Sydney University, Locked Bag 1797, Penrith, NSW, 2751, Australia}
\affiliation{Australia Telescope National Facility, CSIRO, Space and Astronomy, PO Box 76, Epping, NSW 1710, Australia}
\email[Peter. J. Macgregor]{peter.macgregor.astro@gmail.com}

\author{Ray P. Norris}
\affiliation{School of Science, Western Sydney University, Locked Bag 1797, Penrith, NSW, 2751, Australia}
\alsoaffiliation{Australia Telescope National Facility, CSIRO, Space and Astronomy, PO Box 76, Epping, NSW 1710, Australia}

\author{Andrew O'Brien}
\affiliation{Australia Telescope National Facility, CSIRO, Space and Astronomy, PO Box 76, Epping, NSW 1710, Australia}
\alsoaffiliation{School of Science, Western Sydney University, Locked Bag 1797, Penrith, NSW, 2751, Australia}

\author{Mohammad Akhlaghi}
\affiliation[$^{7}$]{Centro de Estudios de F\'isica del Cosmos de Arag\'on (CEFCA), Unidad Asociada al CSIC, Plaza San Juan 1, 44001 Teruel, Spain }

\author{Craig Anderson}
\affiliation{Australia Telescope National Facility, CSIRO, Space and Astronomy, PO Box 76, Epping, NSW 1710, Australia}
\alsoaffiliation{National Radio Astronomy Observatory, PO Box 0, Socorro, NM87801, USA}

\author{Jordan D. Collier}
\affiliation{The Inter-University Institute for Data Intensive Astronomy (IDIA), Department of Astronomy, University of Cape Town, Private Bag X3, Rondebosch, 7701, South Africa}
\alsoaffiliation{School of Science, Western Sydney University, Locked Bag 1797, Penrith, NSW, 2751, Australia}
\alsoaffiliation{Australia Telescope National Facility, CSIRO, Space and Astronomy, PO Box 76, Epping, NSW 1710, Australia}

\author{Evan J. Crawford}
\affiliation{School of Computer, Data, and Mathematical Sciences, Western Sydney University, Locked Bag 1797, Penrith, NSW, 2751, Australia}

\author{Stefan W. Duchesne}
\affiliation{International Centre for Radio Astronomy Research (ICRAR), Curtin University, Bentley, WA 6102, Australia}
\alsoaffiliation{Australia Telescope National Facility, CSIRO, Space and Astronomy, PO Box 1130, Bentley, WA 6151, Australia}

\author{Miroslav D. Filipovi\'c}
\affiliation{School of Science, Western Sydney University, Locked Bag 1797, Penrith, NSW, 2751, Australia}

\author{B\"arbel S. Koribalski}
\affiliation{Australia Telescope National Facility, CSIRO, Space and Astronomy, PO Box 76, Epping, NSW 1710, Australia}
\alsoaffiliation{School of Science, Western Sydney University, Locked Bag 1797, Penrith, NSW, 2751, Australia}

\author{Florian Pacaud}
\affiliation{Argelander-Institut f\"ur Astronomie, Universt{\"a}t Bonn, Auf dem H\"ugel 71, 53121, Bonn, Germany}

\author{Thomas H. Reiprich}
\affiliation{Argelander-Institut f\"ur Astronomie, Universt{\"a}t Bonn, Auf dem H\"ugel 71, 53121, Bonn, Germany}

\author{Christopher J. Riseley}
\affiliation{Dipartimento di Fisica e Astronomia, Universit\`a degli Studi di Bologna, via P. Gobetti 93/2, 40129 Bologna, Italy}
\alsoaffiliation{INAF $-$ Istituto di Radioastronomia, via P. Gobetti 101, 40129 Bologna, Italy}
\alsoaffiliation{Australia Telescope National Facility, CSIRO, Space and Astronomy, PO Box 1130, Bentley, WA 6151, Australia}

\author{Lawrence Rudnick}
\affiliation{University of Minnesota, Minnesota Institute for Astrophysics, 116 Church St. SE, Minneapolis, MN 55455 USA}

\author{Tessa Vernstrom}
\affiliation{ICRAR, The University of Western Australia, 35 Stirling Hw, 6009 Crawley, Australia}
\alsoaffiliation{Australia Telescope National Facility, CSIRO, Space and Astronomy, PO Box 1130, Bentley, WA 6151, Australia}

\author{Andrew. M. Hopkins}
\affiliation{Australian Astronomical Optics, Macquarie University, 105 Delhi Rd, North Ryde, NSW 2113, Australia}

\author{Melanie Johnston-Hollitt}
\affiliation{Curtin Institute for Data Science, Curtin University, Perth, GPO Box U1987, WA 6845, Australia}

\author{Josh Marvil}
\affiliation{National Radio Astronomy Observatory, PO Box 0, Socorro, NM87801, USA}

\author{Matthew Whiting}
\affiliation{Australia Telescope National Facility, CSIRO, Space and Astronomy, PO Box 76, Epping, NSW 1710, Australia}

\author{Steven Tingay}
\affiliation{International Centre for Radio Astronomy Research (ICRAR), Curtin University, Bentley, WA 6102, Australia}

\doi{10.1017/pas.\the\year.xxx}

\received {dd Mmm YYYY}
\revised  {dd Mmm YYYY}
\accepted {dd Mmm YYYY}
\published{dd Mmm YYYY}

 %% First letter not capped

\begin{document}

% PASA uses footnotes, not endnotes. \endnote in this template will behave like \footnote; and \printendnotes will not output anything.
% \printendnotes

%%%%%%%%% ABSTRACT %%%%%%%%%
\begin{abstract}
We present radio observations of the galaxy cluster Abell~S1136 at 888~MHz, using the \acl{ASKAP} radio telescope, as part of the \acl{EMU} Early Science program.
We compare these findings with data from the \acl{MWA}, \acl{XMM}, the \acl{WISE}, the \acl{DSS}, and the \acl{ATCA}. Our analysis shows the X-ray and radio emission in Abell~S1136 are closely aligned and centered on the BCG, while the X-ray temperature profile shows a relaxed cluster with no evidence of a cool core. We find that the diffuse radio emission in the centre of the cluster shows more structure than seen in previous low-resolution observations of this source, which appeared formerly as an amorphous radio blob, similar in appearance to a radio halo; our observations show the diffuse emission in the Abell~S1136 galaxy cluster contains three narrow filamentary structures visible at 888~MHz, between $\sim$80 and 140~kpc in length; however the properties of the diffuse emission do not fully match that of a radio (mini-)halo or (fossil) tailed radio source.
\end{abstract}

%%%%%%%%% INTRODUCTION %%%%%%%%%
\section{Introduction}
\label{section:introduction}

Galaxy clusters are large gravitationally-bound structures, containing hundreds to thousands of galaxies, with typical masses of $10^{14}$ to $10^{15}$\Msun. Approximately 80\% of this mass is Dark Matter, 15$-$17\% is the \ac{ICM}, and only 3$-$5\% is the galaxies that make up the cluster \citep{taxonomy_feretti_2012A&ARv..20...54F}. The \ac{ICM} consists of heated gas situated between the galaxies, including hot X-ray emitting gas at temperatures of $10^7$$-$$10^8$ K \citep[][]{2019A&A...621A..41G}. Galaxy clusters have also been inferred to have magnetic fields of the order of 0.1 to 1~$\mu$G \citep[e.g.][]{2001ApJ...547L.111C, 2002ARA&A..40..319C, 2010A&A...513A..30B, donnert_2018SSRv..214..122D}. 

Diffuse radio emission is often observed in clusters \citep{van-weeren-2019SSRv..215...16V, Paul_2023JApA...44...38P} and is generated by synchrotron radiation, indicating the presence of relativistic particles and magnetic fields \citep{toothbrush-2016ApJ...818..204V,2021map..book....2H}. Because the synchrotron emitting electrons have a limited lifetime, the relativistic particles must either be re-accelerated or created in place \citep{jaffe-1977ApJ...212....1J,2021pma..book.....F}.

The \ac{SED} of the synchrotron radio emission is a function of the ages of the electron populations, the strength of the magnetic fields, and the possible shock-driven re-acceleration from merger events \citep{taxonomy_feretti_2012A&ARv..20...54F, 2014MNRAS.444L..44B,2021map..book....2H}. 

The radio emission from galaxy clusters can be broadly classified into four categories, depending on the location, morphology, and polarisation properties \citep[see][for a recent review]{van-weeren-2019SSRv..215...16V}: (a) radio emission from the constituent galaxies, including bent-tail galaxies; (b) diffuse halos of radio emission (radio halos) broadly positioned on the centre of the cluster;  (c) diffuse elongated radio shocks  typically found on the periphery of the cluster; and (d) fossil radio galaxies.

Radio halos are sources of diffuse radio emission, which are thought to occur more frequently in clusters linked with merger activity, and are associated with the centre of the merging galaxy clusters \citep{brunetti_giant_halo_evolution_2009A&A...507..661B, toothbrush-2016ApJ...818..204V}. The halo spans the region where most of the X-ray emission from the \ac{ICM} is observed, with recent energy injection into the \ac{ICM} and corresponding high X-ray temperature and luminosity \citep[e.g.][]{giovanniniRadiHaloAndRelicCandidatesNRAO-1999NewA....4..141G, cuciti_halos_and_mergers_1_2021A&A...647A..50C, cuciti_halos_and_mergers_2_2021A&A...647A..51C, Paul_2023JApA...44...38P, tumer_halos_and_mergers_2023ApJ...942...79T}. 

Radio halos have traditionally been categorised into two classes based primarily on their sizes: giant radio halos (or just ``radio halos'') with a typical linear size of $\sim$~1~Mpc, and mini-halos which typically have a size of less than 500 kpc. Radio halos have a low surface brightness and a steep radio spectrum, with typical integrated spectral indices of \mbox{$-1.4<\alpha<-1.1$}\footnote{We adopt the convention that the spectral index $\alpha$ is defined by $S_{\nu} \propto \nu^{\alpha}$, where $S_{\nu}$ is the flux density and $v$ is the frequency.}. 

Other important distinctions between mini-halos and giant radio halos include the fact that mini-halos are almost exclusively found in cool-core clusters \citep{van-weeren-2019SSRv..215...16V}, with the mini-halo centered on the radio-loud \ac{BCG}. \citet{gitti_mini_halos_2015aska.confE..76G} show the size of the mini-halo is comparable to that of the central cooling region in the cluster, although recent results with the SKA Pathfinders and Precursors \citep[e.g.][]{Riseley2022-ms1455,Biava2024-cool-core-sample} show diffuse synchrotron emission from mini-halos outside this central cooling region. In addition, \citet{cassano_mini_halo_emission_2008A&A...486L..31C} found that regardless of the origin of the emitting electrons, even though radio halos and mini-halos have similar radio power, a mini-halo radius is typically $\sim$~4 times smaller than a radio halo; implying an $\sim$~50 times larger synchrotron emissivity for the mini-halo; however, some studies \citep[e.g.][]{Riseley2022-ms1455,Riseley2023-a1413} show larger sizes in high sensitivity data, with mini-halos often on size scales comparable to "normal" radio halos.

\citet[][]{botteon_halos_2022A&A...660A..78B} note the difficulty of using a scale size to distinguish between ``normal'' radio halos and mini-halos, given the wide range of cluster masses, so do not distinguish between a mini-halo and a radio halo. As detailed above, recent high sensitivity observations also show ambiguity in the ability to distinguish between these sources based on size. Similarly, we do not make this distinction.

Radio shocks (a.k.a. ``relics'' or ``gischts'') are regions of elongated, diffuse emission that do not have an optical counterpart; typically found around the periphery of a cluster. Their radio spectrum is typically a steep power-law, with spectral indices in the range $-1.5<\alpha<-1.0$, indicating electron cooling in the post-shock region of a shock wave travelling outward \citep[e.g.][]{Giacintucci-2008A&A...486..347G, 2010Sci...330..347V, van-weeren-2011A&A...528A..38V, Stroe-10.1093/mnras/stt2286, Hindson-10.1093/mnras/stu1669,Riseley2022_Abell3266}. Radio shocks have been found to be polarised at the level of $\sim 10-60$\% above 1~GHz in merging galaxy clusters \citep[e.g.][]{1998A&A...332..395E, 2004JKAS...37..323G, 2010Sci...330..347V,Rajpurohit2021-macj0717}. They exhibit strong depolarisation at frequencies $<1$~GHz \citep{2008A&A...489...69B, 2011JApA...32..567P, Ozawa_2015}. 

Fossil radio galaxies (sometimes known as ``phoenices'' or ``revived fossil plasma'' \citep[][]{taxonomy_kempner_2004rcfg.proc..335K}, and confusingly, occasionally as ``relics''), are the remains of dead radio-loud active galactic nuclei (AGN) in which  radio plasma from previous AGN activity is compressed by a  shock wave.  The shock increases both the magnetic field strength inside the plasma and the energy of the plasma particles, resulting in an increase in synchrotron emission. Compared with radio shocks, these sources are generally found at smaller cluster-centric distances \citep[][]{taxonomy_feretti_2012A&ARv..20...54F} and have sizes ranging from $\lesssim 100$~kpc as seen in Abell~2063 and Abell~4038, up to $\sim 350$~kpc as seen in Abell~85. The emission in many cases has a steep curved spectrum \citep[e.g.][]{Riseley2022_Abell3266,2022ApJ...935..168R, Slee_AngResClusters_2001AJ....122.1172S, venturi_shapleyCC_2017Galax...5...16V}, typically steeper than $\alpha \sim -1.5$, characteristic of an old population of electrons. The remnant is found close to the source galaxy, in the inner tens of kpc from the galaxy cluster. When observed with sufficient angular resolution \citep{Slee_AngResClusters_2001AJ....122.1172S} the sources are often filamentary. 

In recent years there has been a dramatic increase in the discovery of \ac{ICM} filaments associated with radio galaxies and galaxy clusters, thanks to high-resolution low frequency sub-GHz observations with new generation telescopes such as the \ac{ASKAP}, MeerKAT, and the \ac{LOFAR} \citep[e.g.][]{Pasini2022_A1550, 2022ApJ...935..168R, velovic_2023MNRAS.523.1933V}. 

Abell~S1136 is located in the Pisces-Cetus supercluster, at RA(J2000) = 23$^{\rm h}$36$^{\rm m}$17.0$^{\rm s}$ DEC(J2000) = $-$31\degr36\arcmin37\arcsec, and was first observed as part of the Abell~Catalogue ``Southern Survey'' in 1988 \citep{Abell_Southern_Extension_1988_PASP_100_1354_10.2307/40679228}. It is reported in the Meta-Catalogue of X-ray Detected Clusters of Galaxies as MCXC J2336.2-3136 \citep[][]{mcxc-2011A&A...534A.109P} with luminosity $L_{500}=0.50\times10^{37}$~W and mass $M_{X,500}=1.29\times10^{14}$~\Msun\, within an $R_{500}=0.75$~Mpc radius. The Abell~S1136 cluster has a spectroscopic redshift of $z=0.0622$ \citep[][]
{s1136_redshift_2dfs_2002MNRAS.329...87D, redshift_2018ApJ...869..145C} on the \ac{BCG} ESO~470$-$20 \citep{bcg_eso-470-20_1989spce.book.....L}.

Observations with the \ac{MWA} \citep[MWA;][]{2013PASA...30....7T, bowman_2013PASA...30...31B} at 168~MHz \citep{oth+16, Duchesne2021} found Abell~S1136 to host steep-spectrum, diffuse emission \citep[][their section 3.2.21]{Duchesne2021}. The observed features of the emission were consistent with a number of cluster radio sources including radio halos, radio relics, and remnant radio galaxies. There is no corresponding 1400~MHz or 843~MHz emission seen in the \ac{NVSS} \citep{nvss-1998AJ....115.1693C} or the \ac{SUMSS} \citep{sumss-bock-1999AJ....117.1578B, sumss-mauch-10.1046/j.1365-8711.2003.06605.x}, implying a steep spectral index.

We present observations of Abell~S1136 using the \ac{ASKAP}
\citep[][]{Science_With_ASKAP_2007PASA...24..174J, johnston08, mcconnell16, askap-hotan}, as part of the Early Science program of the \ac{EMU} survey \citep{EMU_2011PASA...28..215N}. Section~2 describes the observations, which were taken to investigate the structure of the diffuse radio emission centered on the Abell~S1136 galaxy cluster. In Section~3 we present our results, while in Section~4 we discuss the intra-cluster filamentary structure discovered in the Abell~S1136 galaxy cluster, and the implications of these results.

This paper assumes a flat $\Lambda \mathrm{CDM}$ cosmology with 
\\ % typeset cosmology params to new line so they don't hang out of the column
$H_{0}=68~\mathrm{~km}~\mathrm{~s}^{-1}~\mathrm{Mpc}^{-1}, \Omega_{m}=0.3$, and $\Omega_{\Lambda}=0.7$. Using these cosmological parameters, this gives a scale of 1.234~kpc~arcsec$^{-1}$ at a redshift of $z=0.0622$ (the preferred redshift of Abell~S1136).

%%%%%%%%% DATA and OBSERVATIONS %%%%%%%%%
\section{Observations}
\label{section:observations}
%%%%%%%%% 
\subsection{ASKAP Observations}
\label{subsection:ASKAP}

We observed Abell~S1136 on 29$^{\rm th}$ July 2019 with the \ac{ASKAP}, as part of the \ac{EMU} Early Science Broadband (700-1800~MHz) Survey, under Project Code AS034 (Table~\ref{table:s1136-obs}). The observation used a 33 antenna array  with a maximum baseline of 6~km. We used a centre frequency of 888~MHz, and the full instantaneous bandwidth of 288 MHz available at the time, divided into 288~$\times$ 1-MHz channels.

Only 33 of the 36 \ac{ASKAP} antenna were available for this observation. For this observation the 36 \ac{PAF} beams on each antenna were set at a pitch of 0.90 degrees with a 45-degree rotation, and formed into a ``closepack36'' footprint which provides uniform sensitivity without interleaving \citep{norris20}.

The resulting synthesised beam is 12.6\arcsec$\times$ 10.1\arcsec at 888~MHz for {\tt{robust}\,$=0$} weighting. The observation was performed as a single field with 10 hours on source. PKS~1934$-$638 was used as the primary and secondary calibrator, to provide bandpass and gain calibration for imaging. Data calibration and imaging were performed using the ASKAPsoft \citep{askapsoft_2019ascl.soft12003G} data processing pipeline, running at the Pawsey Supercomputing Centre. We produced two sets of images: one at {\tt{robust}\,$=0$} weighting \citep{Briggs1995}, and the other at {\tt{robust}\,$=+2.0$} (Table~\ref{table:s1136-obs}). 

\label{section:DataValidation}

As a check on amplitude calibration across the 888~MHz \ac{ASKAP} observing tile, we cross-matched the \ac{ASKAP} data with the 887.5~MHz  \ac{RACS} \citep{RACS-2020PASA...37...48M} data. We identified 15 unresolved point sources in and around the Abell~S1136 cluster, and measured their flux densities in the \ac{ASKAP} and \ac{RACS} data using the \acl{carta}\footnote{\url{https://cartavis.org/}}\citep[\textsc{carta};][]{carta}. The flux densities are in good agreement, with an average Mean Absolute Deviation error of 5.81\%.

\begin{table}
    \centering
    \caption{Observation and image details of the \ac{ASKAP} data.}
    \label{table:s1136-obs}
        \begin{tabular}{lll}
        \hline
        Pointing Centre RA (J2000)              & 23$^{\rm h}$\,36$^{\rm m}$\,17.00$^{\rm s}$ \\
        Pointing Centre DEC (J2000)             & $-$31\degr34\arcmin37\arcsec \\
        Scheduling Block ID     & 9164 \\
        Antennas                & 33 \\
        Antenna Numbers         & 1, 2, 3, 4, 5, 6, 7, 8 \\
                                & 9, 11, 12, 13, 14,15,16 \\
                                & 17, 18, 19, 20, 21, 22 \\
                                & 24, 25, 26, 27, 28, 29 \\
                                & 31, 32, 33, 34, 35, 36 \\
        Centre Frequency (MHz)  & 888 \\
        Bandwidth (MHz)         & 288 \\
        Channels                & 288 \\
        RMS ($\mu$Jy/beam)      &  \\
        \hspace{3mm} robust~=~0.0 & $\sim$~40 \\
        \hspace{3mm} robust~=~+2.0 & $\sim$~110\\
        PAF Beams               & 36 \\
        Beam (\arcsec$\times$\arcsec) & \\
        \hspace{3mm} robust~=~0.0 & 12.60 $\times$ 10.04 \\
        \hspace{3mm} robust~=~+2.0 & 25.51 $\times$ 21.27\\
        Footprint               & closepack36 \\
        Duration (hr)           & 10.05 \\ \hline
        \end{tabular}
\end{table}

\begin{table}
    \centering
    \caption{Observation and image details of the MWA~2 and \ac{ATCA} data.}
    \label{tab:obs_mwa_and_atca}
    \begin{tabular}{c c c c c}
    \hline
    Band & $\nu_\text{c}$ & $t_\text{obs}$ & $\sigma_\text{rms}$ & Beam \\
    (MHz)  & (MHz) & (min) & (mJy\,beam$^{-1}$) & (\arcsec$\times$\arcsec) \\
    \hline
    \multicolumn{5}{c}{MWA 2} \\ 
    \hline
    72$-$103 & 88 & 62 & 8.6 & $128 \times 126$ \\
    103$-$134 & 118 & 64 & 4.4 & $95.6 \times 94.3$  \\
    139$-$170 & 154 & 78 & 2.6 & $74.0 \times 72.8$ \\
    170$-$200 & 185 & 78 & 2.4 & $62.5 \times 61.6$ \\
    200$-$231 & 216 & 60 & 4.5 & $53.9 \times 53.7$ \\
    \hline
    \multicolumn{5}{c}{ATCA} \\
    \hline
    1076$-$3124 & 2100 & 1207 &  0.012 & $6.31 \times 3.05$ \\
    \hline
    \end{tabular}
\end{table}

\subsection{MWA Observations}
We observed Abell~S1136  with the Murchison Widefield Array \citep[MWA;][]{2013PASA...30....7T} Phase~2 in its `extended' configuration \citep[hereafter MWA~2;][]{wtt+18_v2_2018PASA...35...33W,2019PASA...36...50B}, for the MWA project G0045 (PI Duchesne). This observation was to perform a follow-up survey of the diffuse, non-thermal galaxy cluster emission originally detected by the MWA in its Phase~1. The MWA~2 data cover 72$-$231~MHz, with five 30~MHz observing bands centred on 88~MHz, 118~MHz, 154~MHz, 185~MHz, and 215~MHz. The observing strategy and data reduction process is described in detail by \citet{Duchesne2020}. 
We used {\tt{robust}\,$=0$} weighting and present the observation and image details in Table \ref{tab:obs_mwa_and_atca}.

%%%%%%%%% 
\subsection{Australia Telescope Compact Array Observations}
\label{sec_data_atca}

We observed Abell~S1136 with the Australia Telescope Compact Array \citep[ATCA;][]{atca-frater, atca-nelson, wilson_cabb_2011MNRAS.416..832W} over two sessions and using two compatible array configurations\footnote{\url{https://www.narrabri.atnf.csiro.au/operations/array_configurations/configurations.html}}, at 1100$-$3100~MHz (the weighted central frequency after the removal of radio frequency interference is 2120~MHz). The first observation was with a 6A array on the 29$^{\rm th}$~April~2016 (PI Duchesne) under project code CX356. This array consists of baselines ranging from 337~m to 6~km, with only 3 of the 15 baselines shorter than 1087~m (337~m, 628~m, and 872~m). The second observation was on 15$^{\rm th}$~November~2020 (PI Macgregor) using a 6B array, under project code CX476. This array consists of baselines ranging from 214~m to 6~km, with only 4 of the 15 baselines shorter than 1270~m (214~m, 536~m, 750~m, and 949~m). This gives a predicted RMS (1$\sigma$) noise level for a 12 hour observation of 9~\ujybm\ which is close to what we measure in our image as 1$\sigma$ = 12~\ujybm.

The data were calibrated using \textsc{miriad} \citep{miriad_1995ASPC...77..433S}. The primary calibrator for each observation was PKS~1934$-$638, and gain calibrators PKS~2313$-$340 (CX356) and PKS~2337$-$334 (CX476). We imaged the observation using \textsc{mfclean}, with the bandwidth left at 2000~MHz, and {\tt{robust}\,$=0$}. Three rounds of phase-only self-calibration were performed, and images were cleaned using \textsc{mfclean} down to 3 times the RMS noise. The \ac{ATCA} observation and image details are shown in Table~\ref{tab:obs_mwa_and_atca}.

The primary beam of the \ac{ATCA} at 2100~MHz centre frequency is 23.1\arcmin\ \ac{FWHM}, which is comparable with the size of the images shown in this paper. The Compact Array Broadband Backend (CABB) \citep{wilson_cabb_2011MNRAS.416..832W} provides a 2~GHz bandwidth when observing at 2100~MHz; this gives a \ac{FWHM} of 45.07\arcmin at 3214~MHz, and  a 15.52\arcmin \ac{FWHM} at 1076~MHz. Although the \ac{ATCA} data are corrected for the primary beam shape, the flux densities of sources far (i.e. $\gtrsim$~11.5\arcmin\ radius / 23\arcmin\ diameter) from the pointing centre (at RA(J2000) = 23$^{\rm h}$\,36$^{\rm m}$\,17.23$^{\rm s}$ DEC(J2000) = $-$31\degr\,36\arcmin\,14.\arcsec22, which is close to the \ac{BCG}), may be unreliable as, at points beyond the \ac{FWHM} the effective observing frequency varies with distance from the pointing centre \citep{2017isra.book.....T}, and the errors on the flux estimates become larger.

Additionally, as the \ac{ATCA} observations had limited short spacing information, they were insensitive to the extended diffuse emission; so we do not attempt to derive spectral indices of the diffuse emission from the \ac{ATCA} data, but use the \ac{ATCA} observations only for compact sources and narrow filaments.

%%%%%%%%% RESULTS and ANALYSIS %%%%%%%%%
\section{Results}
\label{section:results}
%%%%%%%%% 
\subsection{Filamentary Structures}
\label{subsection:Filamentary_sub_Structure}

\begin{figure*}
\centering
    %%%%%%% A
    \begin{subfigure}[t]{0.95\textwidth}
        \centering
        \includegraphics[width=0.65\textwidth]{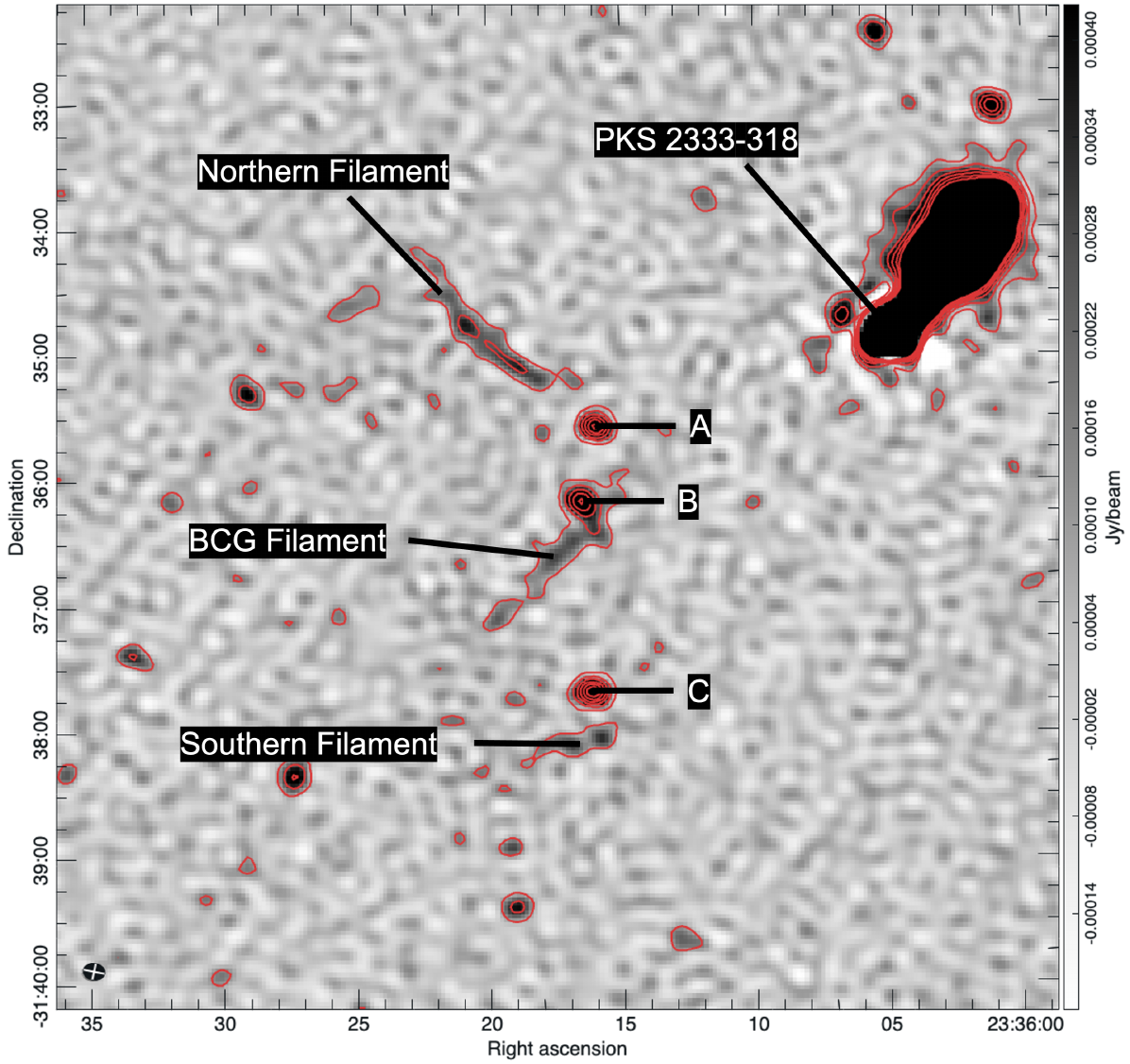}
        \caption{}
        \label{fig:1a_r0_contours}
    \end{subfigure}
    %%%%%%% B
    \\
    \begin{subfigure}[t]{0.95\textwidth}
        \centering
        \includegraphics[width=0.65\textwidth]{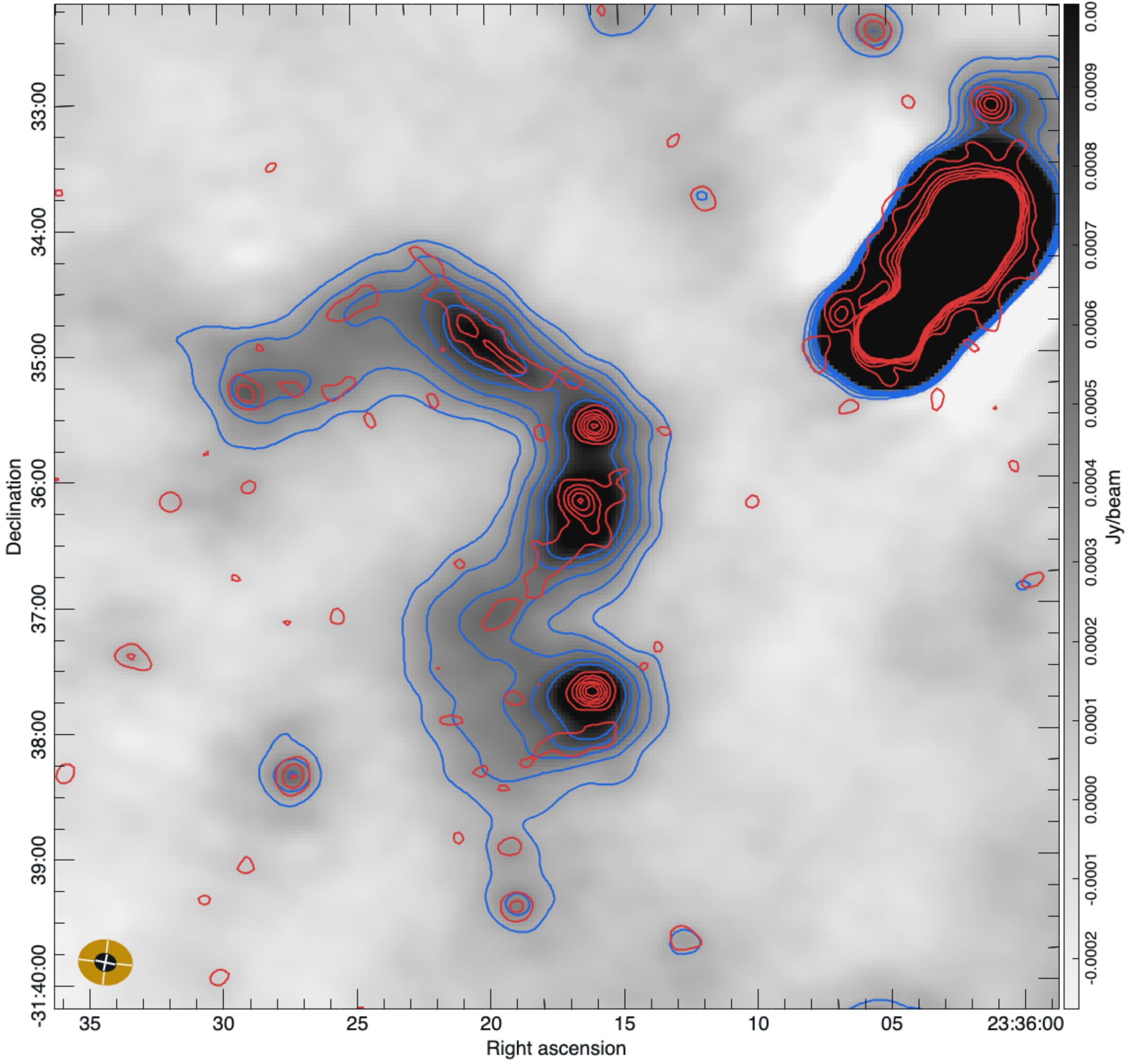}
        \caption{}
        \label{fig:1b_r2_contours}
    \end{subfigure}
    %%%%%%% caption
    \caption{ASKAP 888~MHz radio continuum images of the galaxy cluster Abell~S1136. (a) The high resolution 12.6\arcsec$\times$10.0\arcsec {\tt{robust}\,$=-0.5$} image shows the sources associated with the newly discovered filaments (detailed in Table \ref{table:optical_and_ir_properties}): (A) WISEA J233615.95$-$313534.4 (B) WISEA J233616.55$-$313609.3 and (C) WISEA J233616.10$-$313741.1 labelled together with the radio bright head-tail galaxy PKS~2333$-$318. The contour levels in red start at 2$\sigma$ = 80~\ujybm and scale by a factor of $\sqrt{2}$ to 750~\ujybm.  (b) shows the 25.5\arcsec$\times$ 21.3\arcsec low resolution {\tt{robust}\,$=+2.0$} image of the same sources, plus the diffuse cluster emission. The contour levels in blue start at 2$\sigma$ = 220 \ujybm and scale by a factor of $\sqrt{2}$ to 750~\ujybm. The contour levels in red are the same as Figure \ref{fig:1a_r0_contours}. The diffuse emission component, excluding point sources, was measured at 4.88~mJy$\pm$~0.50; see \S~\ref{subsection_noisechisel} and Table~\ref{table:flux_properties}.}
    \label{fig:s1136_filaments} 
\end{figure*}

With the higher sensitivity and resolution of the \ac{ASKAP} data, the steep-spectrum and diffuse cluster emission previously identified by \citet{Duchesne2021}, can be seen in Figures \ref{fig:1a_r0_contours} and \ref{fig:1b_r2_contours} to break up into several more compact structures. Radio continuum emission detected in these images are divided into regions of diffuse emission with filamentary structure. We identify three filamentary structures:

\begin{enumerate}
    \item The ``Northern Filament'', which appears to extend north-east from WISEA~J233615.95$-$313534.4, the source labelled ``A''. This filament is $\sim~140$~kpc in length (assuming a distance to Abell~S1136 of 250~Mpc), and appears detached from  source A.\\
    
    \item   The ``\ac{BCG} Filament'', which appears to extend south-east from WISEA~J233616.55$-$313609.3 / ESO~470$-$20, which is the Abell~S1136 Brightest Cluster Galaxy \citep{bcg_eso-470-20_1989spce.book.....L}  labelled ``B''. This filament is $\sim 140$~kpc long, and appears to be in projection, or to be attached to, source ``B''. \\
    
    \item The ``Southern Filament'', which  appears to extend nearly horizontally under WISEA~J233616.10$-$313741.1, the source labelled ``C''. The filament appears to be oriented tangentially to the cluster. This filament is the shortest at $\sim 80$~kpc in length. 
\end{enumerate}

Sources A and B are cluster radio galaxies, while C is a likely background source, as described below.

Figure \ref{fig:1a_r0_contours} shows that the Abell~S1136 galaxy cluster has multiple \ac{AGN} located near the centre, with a radio-loud \ac{BCG}. In Figure \ref{fig:1b_r2_contours}, the large-scale diffuse radio emission is largely symmetric around the \ac{BCG}, with a morphology that resembles a \ac{WAT} source. However, Figure \ref{fig:1a_r0_contours} shows that the high-resolution image does not resemble a \ac{WAT}, in that it consists of two relatively straight filaments that do not curve towards each other, unlike the jets of most \acp{WAT}. Furthermore, the northern filament is not connected to the BCG filemant or to the BCG. It might be argued that the low-resolution 154~MHz image in Fig.~2(c) shows two radio lobes, but at higher resolution (e.g. Fig.~2(f)) these break up into a combination of diffuse emission and compact sources, in which the filaments do not point to the patches of diffuse emission. Therefore, while we cannot rule out that the Northern and \ac{BCG} filaments are sections of a \ac{WAT} associated with the \ac{BCG}, this would be a very unusual \ac{WAT}.

%%%%%%%%% 
\subsection{Abell~S1136 flux density measurements}
\label{subsection:flux-density}

We imaged the Abell~S1136 galaxy cluster and its filamentary and diffuse components (see Table~\ref{table:flux_properties}) using data from the MWA Phase~II at 154~MHz, 185~MHz, and 216~MHz, \ac{ASKAP} at 888~MHz and \ac{ATCA} at 2100~MHz, as shown in Figure~\ref{fig:detection}. 

For each of the diffuse regions and filaments measured in the \ac{ASKAP} image, we used NoiseChisel (see below) to measure the integrated flux density of the same regions in the \ac{ATCA} data at 2100~MHz, after images were regridded and convolved to the \ac{ASKAP} beam size using \textsc{miriad} \citep{miriad_1995ASPC...77..433S}.

%%%%%%%%% 
\subsubsection{Properties of the diffuse emission and filaments}
\label{subsection_noisechisel}

To detect and measure diffuse emission, we used the \textsc{NoiseChisel}\footnote{The following NoiseChisel options were changed compared to the default (using Gnuastro version \maGnuastroVersion): because radio data are already heavily smoothed, convolution was disabled with \textsc{\maNoiseChiselkernel}, the quantile-threshold was increased with \textsc{\maNoiseChiselqthresh} and the growth-threshold was decreased to \textsc{\maNoiseChiseldetgrowquant}. Furthermore, to increase purity, we used \textsc{\maNoiseChiselsnquant}.} and \textsc{MakeCatalog} tools, which are part of the GNU Astronomy Utilities \citep[Gnuastro; ][]{akhlaghi15,akhlaghi19}. NoiseChisel uses very low thresholds to nonparametrically detect  extended regions of elevated brightness, to very low signal-to-noise (S/N) ratios \citep{akhlaghi21}.

To measure the uncertainty of this integrated flux density, we used MakeCatalog's \textsc{-{}-upperlimit}: the footprint of the region was randomly placed in parts of the image with similar noise properties\footnote{The input images in all bands are much larger than the small crop shown in Figure \ref{fig:detection}, allowing good sampling over the noise.}, discarding any position where the displaced footprint overlaps with a detection, and the sum of pixel values was measured. The process was terminated after $\maUpNum$ successful measurements (\textsc{-{}-upnum=10000}). The sigma-clipped median ($m_r$) and standard deviation ($\sigma_r$) of this random distribution were calculated and compared with the total sum within the region ($s$), giving an estimate of significance of $(s-m_r)/\sigma_r$.

Figure \ref{fig:detection}(a) shows the NoiseChisel detection of diffuse emission in Abell~S1136 as the green-shaded pixels. Figure \ref{fig:detection}(b) shows other likely detections in the image at the same level, as black pixels, showing  that similar extended regions of diffuse emission are not found elsewhere in the image.

Using the 888~MHz \ac{ASKAP} {\tt{robust}\,$=0$} image, shown in Figure~\ref{fig:detection}(a),
we used NoiseChisel to measure the flux density of the three filaments above a surface brightness threshold of 100 \ujybm (shown as a purple outline in the green emission in Figure~\ref{fig:detection}a). Similarly, we measured the flux density of the diffuse emission (shown in green) using only NoiseChisel-detected pixels above this threshold. The results are listed in Table~\ref{table:flux_properties}. In each case, uncertainties were estimated as described above.

Using the uncertainty estimation technique described above, we obtain a significance for this detection of $\maDiffuseAllSig\sigma$. To confirm that this high fraction is not due to a concentrated region (that may have been missed in the masking) and is mostly uniform, we manually divided the diffuse region into two, shown as blue and red in Figure \ref{fig:detection}(b). We found a significance of $\maDiffuseSigmaOne\sigma$ and $\maDiffuseSigmaTwo\sigma$ for the blue and red regions respectively. The results of all measurements are shown in Table~\ref{table:flux_properties}. 

%%%%%%%%% 
\subsubsection{Total flux density}

\begin{figure*}
  \includegraphics[width=\textwidth]{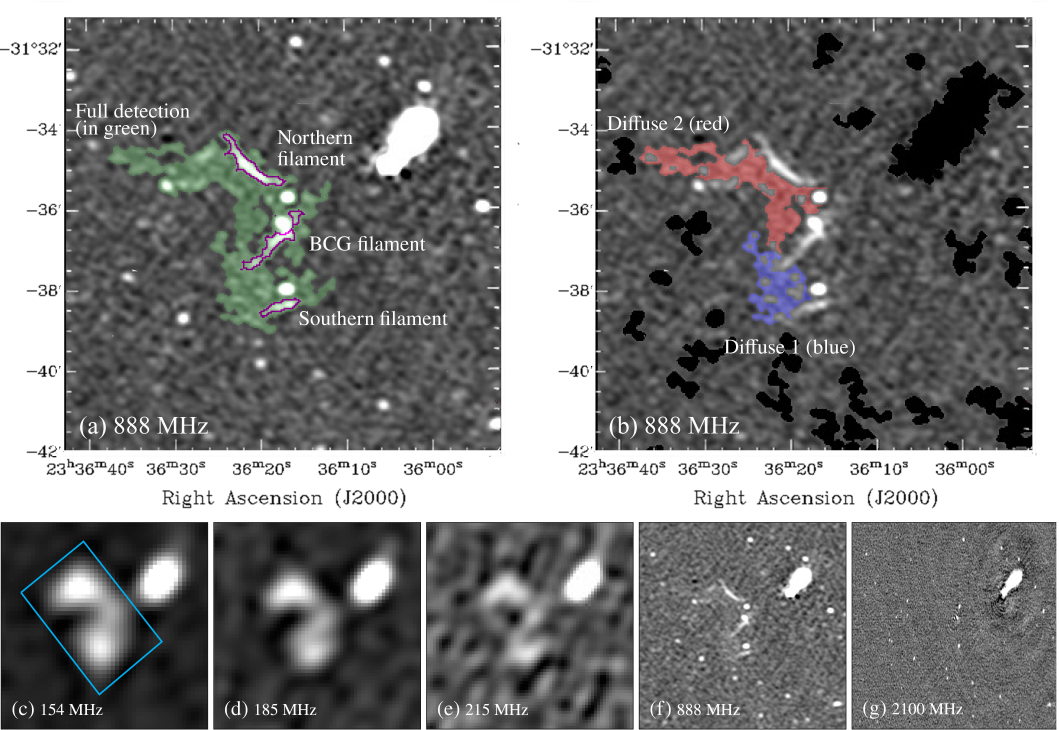}
  \vspace{-5mm}
  \caption{Detections in the galaxy cluster Abell~S1136.
    (a) \ac{ASKAP} 888~MHz (12.6\arcsec$\times$10.0\arcsec). The green region shows the \textsf{NoiseChisel}-detected region ($\maDiffuseAllSig\sigma$).
    The purple outline regions show the pixels within each filament using a 100 \ujybm cut-off.
    (b) \ac{ASKAP} 888~MHz (12.6$\times10.0$\,arcsec$^2$). The diffuse region is manually divided into the  blue and red regions (Diffuse 1 and Diffuse 2) to measure the significance of the detection in them ($\maDiffuseSigmaOne\sigma$ and $\maDiffuseSigmaTwo\sigma$ respectively). Other putative detections in the image at the same level are shown as black pixels, demonstrating that there are no other regions in the image with similar extended and contiguous emission. 
     (c) to (g) show the images obtained by each telescope at the frequency shown. (c),(d),(e) are from MWA~2 with resolutions at 74.0\arcsec$\times$72.8\arcsec, 62.5\arcsec$\times$61.6\arcsec, and 53.9\arcsec$\times$53.7\arcsec; (f) is from \ac{ASKAP} (12.6\arcsec$\times$10.0\arcsec, and (g) is from \ac{ATCA} (6.31\arcsec$\times$3.05\arcsec). The blue box in figure (c) is the region in which the total emission is measured, in order to exclude any contribution from PKS~2333$-$318.}
  \label{fig:detection}
\end{figure*}

The total integrated flux density measurements for the cluster were obtained using images at all frequency bands, shown in Figures~\ref{fig:detection}(c$-$g). In the lowest resolution MWA~2 image at 154~MHz, the galaxy cluster Abell~S1136 and the head-tail radio galaxy PKS~2333$-$318 are the only significant radio detections. To measure the integrated spectral index of Abell~S1136, we defined a region, shown as a cyan box in Figure~\ref{fig:detection}(c), to exclude the emission from PKS~2333$-$318 and contain only the emission from Abell~S1136. For the other frequencies (185, 215, 888, and 2100~MHz) we measured the integrated flux density within the same box. The resulting integrated flux densities and spectral index are shown in Table~\ref{table:flux_properties}.

\begin{table*}
\centering
    \caption{Integrated flux densities and spectral indices of the Abell~S1136 emission components. The flux densities of the filaments and diffuse regions (and their one-sigma uncertainties) were measured using NoiseChisel, as described in Section \ref{subsection:flux-density}. These errors do not account for systematic errors (generally 10\%).
    No spectral index is given for the filaments as the \ac{ATCA} measurements may be affected by missing short spacings.}
    \label{table:flux_properties}
    \renewcommand{\arraystretch}{1.45}
    \begin{tabular}{@{}lcccccc@{}}
    \hline
    Source & $S_{\rm 154\,MHz}$ & $S_{\rm 185}$MHz & $S_{\rm 215}$MHz & $S_{\rm 888}$MHz & $S_{\rm 2100}$MHz 
    & Spectral index ($\alpha$) \\
    & \multicolumn{5}{c}{ (mJy)} \\
    \hline
    
    Northern Filament           & $-$  & $-$ & $-$      & $2.43\pm\maCatHiResFileeeNe$  & $\maCatHiResFiltozzN\pm\maCatHiResFiltozzNe$ & 
    $-$ \\
    
    BCG Filament                & $-$  & $-$ & $-$  & $1.94\pm\maCatHiResFileeeBe$  & $\maCatHiResFiltozzB\pm\maCatHiResFiltozzBe$ & 
    $-$ \\
    
    Southern Filament           & $-$  & $-$ & $-$  & $0.92\pm\maCatHiResFileeeSe$  & $\maCatHiResFiltozzS\pm\maCatHiResFiltozzSe$ &  
    $-$ \\
    
    ``Green'' Diffuse Region $^\text{a}$ & $-$ & $-$ & $-$ & $\maCatHiResDifeee\pm\maCatHiResDifeeee$    &  
    $-$ & 
    $-$    \\
    
    Box Total $^\text{b}$  & $455.7\pm6.5$ & $260.3\pm8.5$ & $114.4\pm16.6$ & $\maCatLowReseee\pm1.3$
    &
    $-$ & $
    -1.68^{+\maCatLowResSpecIndP}_{-\maCatLowResSpecIndN}$ $^\text{c}$  \\
     
    \hline
\end{tabular} \\
\begin{flushleft}
    {\scriptsize
  $^\text{a}$ The green diffuse region is shown in Figure \ref{fig:detection}(a). \\
  $^\text{b}$ The box totals have been calculated within the cyan box shown in Figure \ref{fig:detection}(c). This includes both diffuse emission and discrete sources. \\
  $^\text{c}$ The spectral index measurement also includes both diffuse and discrete sources, and as discussed in the text is probably flatter than the emission from the diffuse emission alone. }
\end{flushleft}

\renewcommand{\arraystretch}{1} 
\end{table*}

%%%%%%%%%
\subsection{Spectral Index}
\label{subsection:Spectral-Index}

%%%%%%%%%
\label{subsection:Spectral_Index_Map}

A low-resolution spectral index map was  produced using  
\\ % typeset {robust} to new line so it doesn't hang out of the column
{\tt{robust}\,$=0$} images from MWA~2 at 154, 185, and 215~MHz, and \ac{ASKAP} at 888~MHz. The data were regridded to the finest pixel grid (ASKAP 888~MHz 12.6\arcsec$\times$10.0\arcsec) and convolved to the largest beam (MWA~2 154~MHz 73.95\arcsec$\times$72.85\arcsec). The images were then stacked together, with the individual pixels in the stacked images processed through a simple weighted linear regression algorithm, with the slope of the best line of fit saved to a new image and output as a fits file. This image is shown in Figure \ref{fig:spix-map}. It is important to note that the resolution of the low frequency measurements does not allow the separation of the spectral index of the diffuse emission from that of the compact sources. 

\begin{figure}
\centering
    \hspace{-0.8cm}
    \includegraphics[trim=0 0 0 0, width=\textwidth]{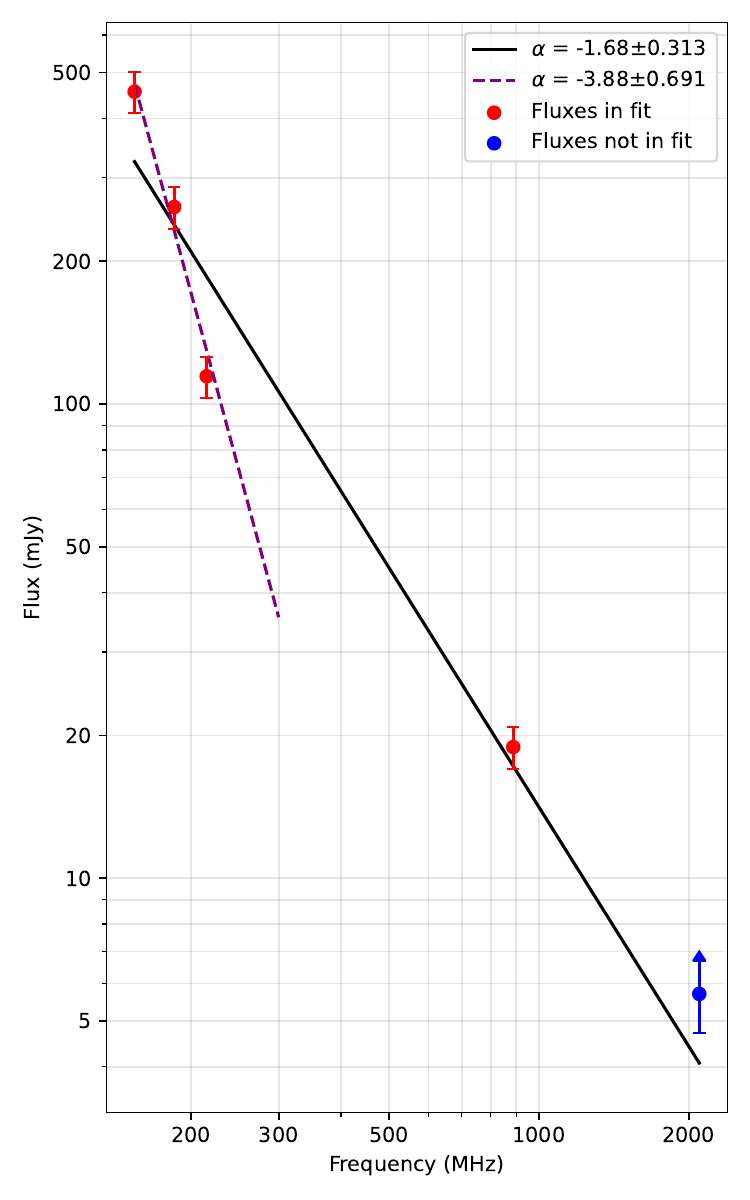}
    \caption[]{The spectral index measurements using the integrated flux of the diffuse emission and compact components in the Abell S1136 galaxy cluster. The flux measurements from Table \ref{table:flux_properties} have been weighted and plotted with error bars derived from a quadrature sum of estimated 10\% systematic errors plus the errors from Table \ref{table:flux_properties}. The solid black line shows the fit between \ac{MWA} 154, 185, and 215~MHz, and \ac{ASKAP} 888~MHz, with a spectral index of $\alpha\sim-$1.68. The \ac{ATCA} 2100~MHz spectral index is plotted in blue and shown here as a lower limit. As described in Section \ref{sec_data_atca}, the \ac{ATCA} flux is not included in the fit as the short spacings are not sensitive to the extended diffuse emission in Abell~S1136. There is indication of a very steep spectrum component at low frequencies in the three \ac{MWA} data points, with a spectral index of $\alpha\sim-$3.88. This is consistent with a steep-spectrum diffuse component which dominates at low frequencies, but is faint at high frequencies. Including both compact sources and diffuse emission means there are a number of flatter-spectrum compact sources which dominate the total flux density, helping to flatten the spectral index; particularly at higher frequencies.}
    \label{fig_spix_curvature}
\end{figure}

\begin{figure*}
\centering
    \includegraphics[trim=0 0 0 0, width=0.8\textwidth]{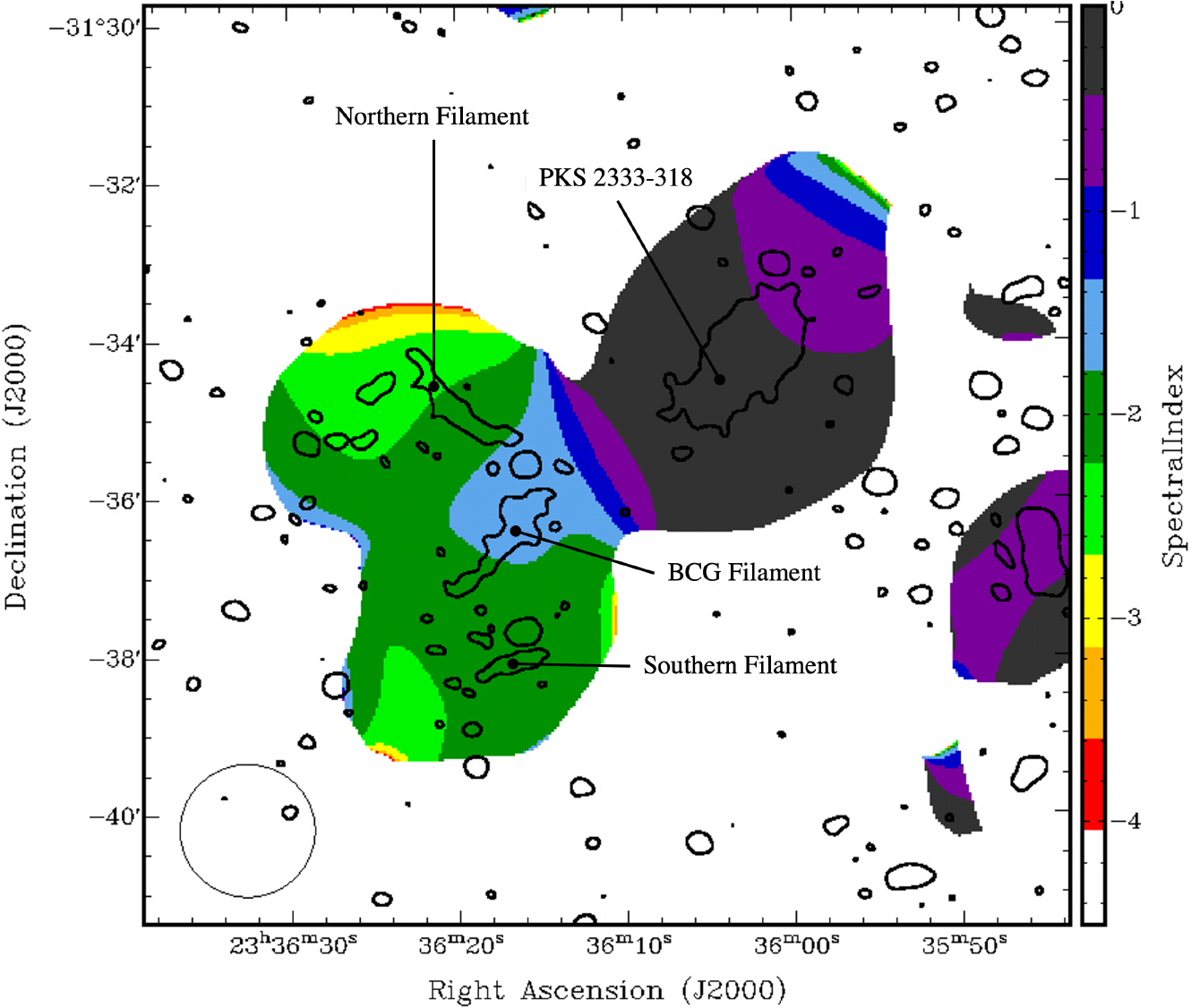}
    \caption[]{The spectral index map of the galaxy cluster Abell~S1136. The map was created using images from MWA~2 @ 154~MHz (74.0\arcsec$\times$72.8\arcsec), 185 (62.5\arcsec$\times$61.6\arcsec) MHz, and 215~MHz (53.9\arcsec$\times$53.7\arcsec), and \ac{ASKAP} 888~MHz (12.6\arcsec$\times$10.0\arcsec). Contours in black are at 100 \ujybm\ from the \ac{ASKAP} 888~MHz image. The data were regridded to the finest pixel grid and convolved to the largest beam. The MWA~2 beam is shown in the bottom left corner.}
    \label{fig:spix-map}
\end{figure*}

The \ac{MWA} integrated flux measurements at 154, 185, and 215~MHz (see Table \ref{table:flux_properties} and Figure \ref{fig_spix_curvature}), show a clear indication of a very steep spectrum $\alpha\sim-$3.88 component at low frequencies. Inclusion of the 888~MHz \ac{ASKAP} integrated flux flattens the overall measured spectral index to $\alpha\sim-$1.68, shown by the solid black line in Figure \ref{fig_spix_curvature}. This is consistent with a steep-spectrum diffuse component which dominates at low frequencies but is faint at high frequencies, combined with  a number of flatter-spectrum compact sources which dominate the total flux density at high frequencies, helping to flatten the spectral index fit. It is therefore difficult to determine if the diffuse emission alone is consistent with a  power-law spectrum  \citep[][]{van-weeren-2019SSRv..215...16V}.

%%%%%%%%%
\subsection{X-ray Data}
\label{subsection:xray}

\begin{figure*}
\centering
    \includegraphics[width=0.8\textwidth]{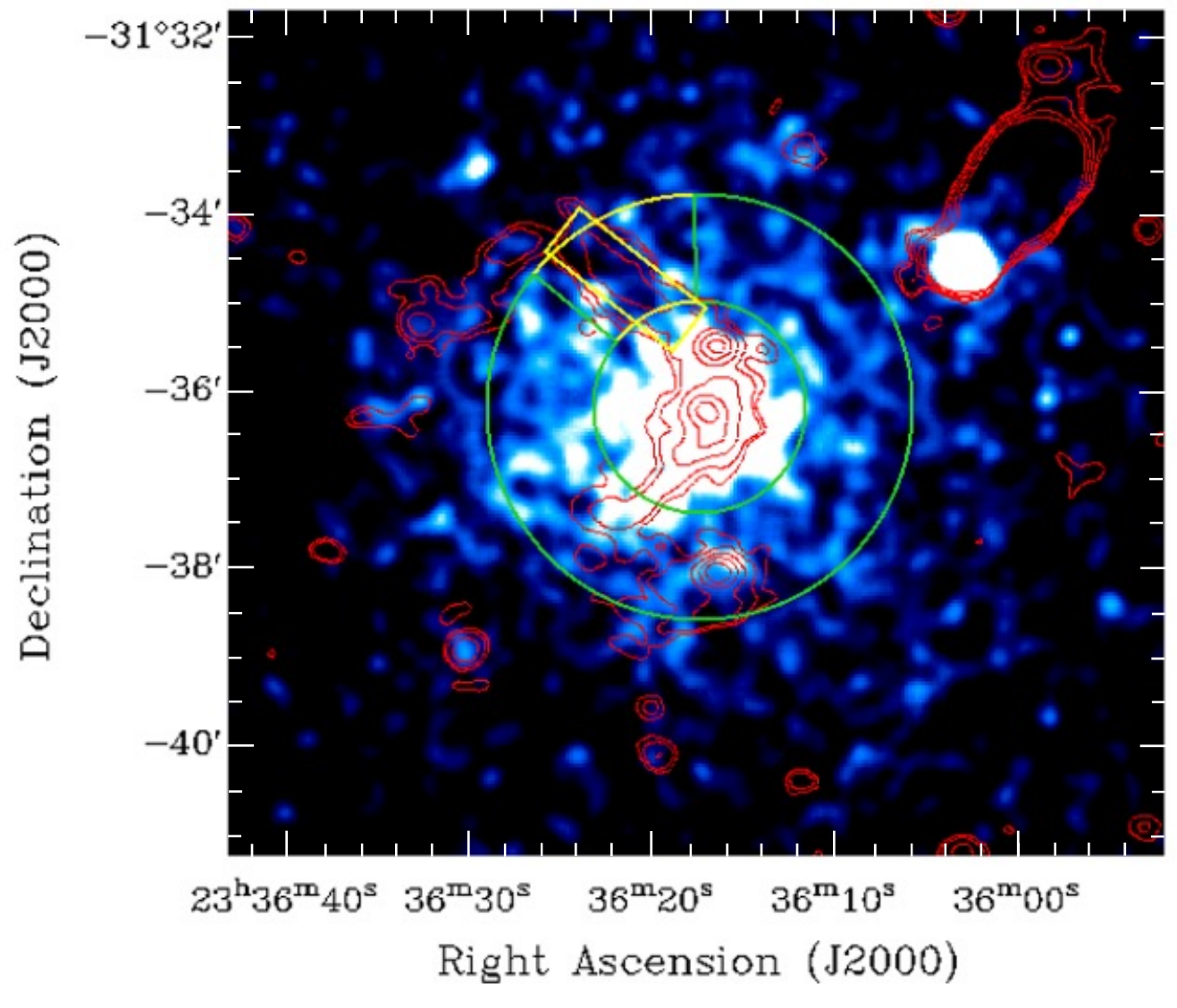}
    \caption[]{XMM-Newton X-ray image (0.5$-$2~keV) of the galaxy cluster Abell~S1136 convolved to \ac{ASKAP} 888~MHz low resolution image (Figure \ref{fig:1b_r2_contours}; 25.51\arcsec$\times$21.27\arcsec), showing an X-ray peak located at RA(J2000) = 23$^{\rm h}$36$^{\rm m}$16.54$^{\rm s}$ DEC(J2000) = $-$31\degr36\arcmin09.\arcsec5, coincident with the radio emitting \ac{BCG}~ESO~470-G020. \ac{ASKAP} contours (red) are at 0.1, 0.15, 0.3, 0.6, and 1.0~mJy~beam$^{-1}$. PKS~2333$-$318 (to the north-west of the cluster centre) is well within the projected extent of the Abell~S1136 galaxy cluster \ac{ICM}. The yellow rectangle (30\arcsec wide) indicates a possible ``channel'' of marginal ($\sim 4.3 \sigma$) significance. The statistical significance of the channel  was calculated using emission between 1 and 2\arcmin of the cluster centre (indicated by the inner and outer green rings) but excluding the region delineated by the two radial green lines, as described in Section \ref{channel}.}
    \label{fig:askap_and_xmm}
\end{figure*}

Figure \ref{fig:askap_and_xmm} shows the X-ray image of Abell~S1136 from observations with the XMM-Newton X-ray telescope (Id 0765041001) obtained as part of the \ac{eeHIFLUGCS} follow-up project \citetext{\citealt{eeHIFLUGCS_2017xru..conf..189R}; \citealt{2019A&A...626A..48R}; \citealt{migkas_xray_2020A&A...636A..15M}; Pacaud et al. in prep.}. 

The data were processed and calibrated using the standard XMM-SAS processing tools (version 18.0.0). The calibrated event-lists were filtered for possible particle flare contamination based on a Poisson distribution fit to the [0.3-10.0] keV light curve values in time bins of 52 / 26s (resp. for the MOS / PN detectors). Time bins with count arrival rates outside the 3$\sigma$ central range of the Poisson distribution were excluded as contaminated. This resulted in the exclusion of only 9\% of the total exposure time. The instrumental background for each \ac{XMM} / EPIC exposure was modelled based on Filter Wheel Closed observations rescaled using the signal observed in the unexposed CCD corners. Images were extracted in the [0.5-2.0] keV band, resulting in the instrumental background subtracted, exposure corrected surface brightness map displayed in Figure \ref{fig:askap_and_xmm}. The morphology of the X-ray map confirms the emission peak to be coincident with the \ac{BCG} defined at the coordinates in the \ac{ASKAP} observations. The \ac{BCG} is often associated with the geometric and kinematic galaxy centre, as well as the peak of the cluster X-ray emission \citep{K-BAND_PropertiesOfGalaxyClusters_2004ApJ...617..879L,cool_core_temp_drop_2010A&A...513A..37H}. 

Some features are further visible in the map that connects to the radio emission; (a) a possible excess coincident with the \ac{BCG} filament, seen as a bright ridge which extends along the \ac{BCG} radio emission, and (b) a possible ``channel'' of missing thermal plasma at the position of the northern filament, shown by the yellow rectangle in Figure \ref{fig:askap_and_xmm}.  We discuss this further in Section \ref{channel}.

%%%%%%%%%
\subsubsection{X-ray Temperature and Spectral Profile Analysis}
\label{section_xmm_askap_temperature_Analysis}

To perform a more detailed spectroscopic analysis of the X-ray data, and measure other X-ray properties, we created a mask of the X-ray point sources to be excluded from the analysis. We first modelled the sky background from the outer regions of the pointing assuming that it consists of a contribution from the unresolved AGN population (an absorbed power-law spectrum with an X-ray spectral index of 1.46), the emission from the Milky Way halo (an absorbed \ac{APEC} thermal model with solar metal abundance and a fitted temperature of $\sim$0.2\,keV) and that of the local hot bubble (again absorbed solar metalicity \ac{APEC}, with a temperature $\sim$0.1\,keV). Our spectral modelling relied on version 3.0.9 of the \ac{APEC} model and a fixed hydrogen column density of $1.23\times10^{20}$\,cm$^{-2}$ derived from the Leiden/Argentine/Bonn (LAB)  survey \citep[][]{healpix_2005ApJ...622..759G, LAB_survey_2005A&A...440..775K, land_2007PhRvD..76h7301L}. Then we proceeded with extracting and modelling spectra in regions of interest, adding a further cluster \ac{APEC} component to the area-rescaled sky background.

\begin{figure*}
    %%%%%%% A
    \hspace*{-1cm} 
    \begin{subfigure}[]{0.52\linewidth}
        \centering
        \includegraphics[width=\linewidth]{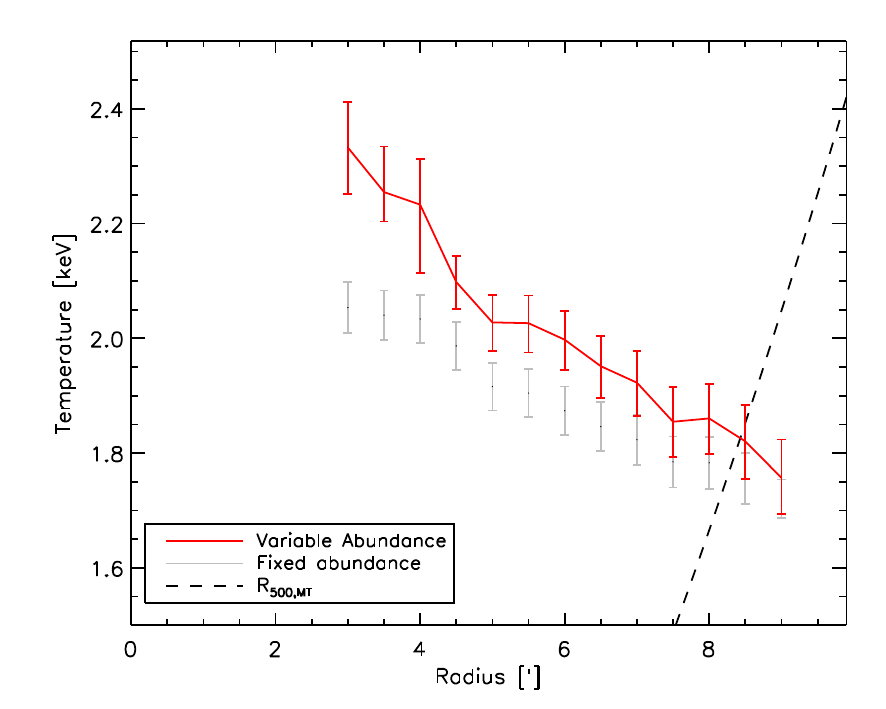}
        \caption{Abell~S1136 X-ray Average Cluster Temperature}
        \label{fig:xray_avg_temp}
    \end{subfigure}
    %%%%%%% B
    \begin{subfigure}[]{0.52\linewidth}
        \centering
        \includegraphics[width=\linewidth]{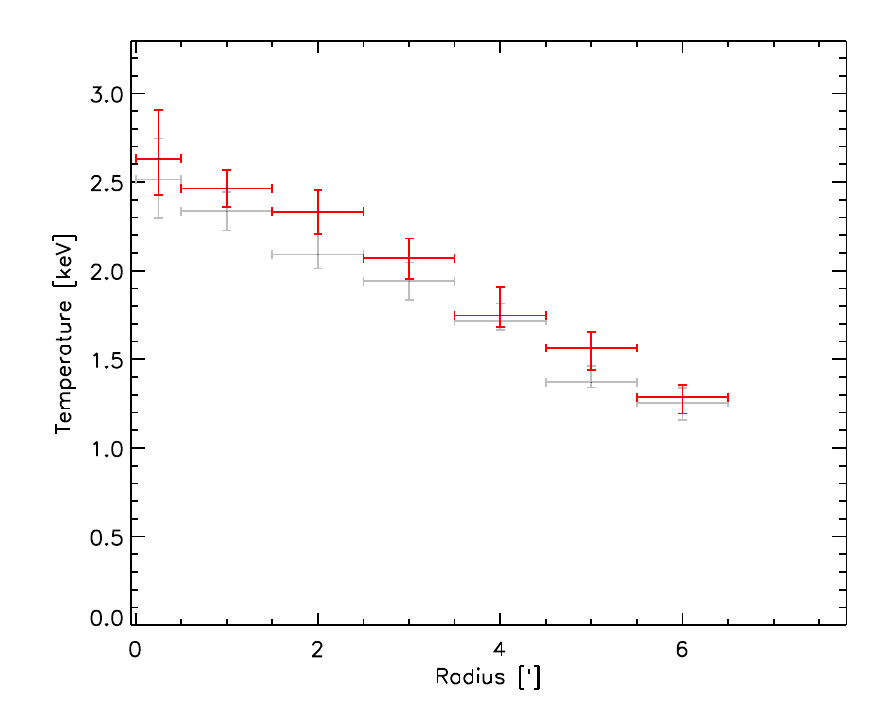}
        \caption{Abell~S1136 X-ray Temperature Profile}
        \label{fig:xray_temp_profile}
    \end{subfigure}
    %%%%%%% caption
    \caption{The X-ray temperature profiles of the Abell~S1136 galaxy cluster; in both Figures the red line is the result leaving the metal abundance free, the grey line is for a fixed value of 0.3 solar. The location on the x-axis on both plots is the radial distance from the \ac{BCG} at which the measurement is taken. Figure \ref{fig:xray_avg_temp} shows the average cluster temperature (expressed as kT), estimated in apertures of increasing size from the cluster centre. The dashed line shows the average $R_{500}-T$ relation for a cluster at $z = 0.062$, based on the galaxy cluster mass-temperature (M$-$T) relation of \citet{arnaud_2005A&A...441..893A}. The average temperature of the Abell~S1136 galaxy cluster at 8.4\arcmin (the $R_{500}$ radius) is $\sim$~1.82~keV, shown by the intersection of the dashed and red lines in Figure \ref{fig:xray_avg_temp}. Figure \ref{fig:xray_temp_profile} shows the Abell~S1136 X-ray temperature profile, with the cluster temperature plotted against radial distance in arcminutes from the cluster centre. The temperature profile appears linear from the centre to the edge, with no drop in temperature towards the cluster centre.}
    \label{fig:xray_temperature_profiles} 
\end{figure*}

The red and grey points in Figure~\ref{fig:xray_avg_temp} show the average cluster temperature estimated in apertures of increasing size. The dashed line shows the average $R_{500}-T$ relation for a cluster at $z = 0.062$, based on the galaxy cluster mass-temperature (M$-$T) relation of \citet{arnaud_2005A&A...441..893A}. The average temperature of the Abell~S1136 galaxy cluster at the $R_{500}$ radius of 8.4\arcmin is $\sim$~1.82~keV, shown by the intersection of the dashed and red lines in Figure \ref{fig:xray_avg_temp}. This enables us to estimate an approximate cluster $R_{500} / M_{500}$ (the halo mass $M_{500}$ out to a radius $R_{500}$ within which the total mass density is 500 times the critical density of the Universe). Other cluster average properties are estimated from a direct fit to spectra extracted within $R_{500}$ and shown in Table \ref{table:abell_xray_luminosity}. 

We note that the values shown in Table 5 differ slightly from the earlier ROSAT measurements \citep{migkas_xray_2020A&A...636A..15M, mcxc-2011A&A...534A.109P} because (a) the ROSAT data include contributions from AGN such as PKS~2333$-$318, and (b) the ROSAT temperature is based on data in the radius range $0.2-0.5 \times R_{500}$ (i.e. excluding the central 2\arcmin)  whereas ours is based on all the emission within $R_{500}$.

\begin{table}
\centering
\caption[]{The Abell~S1136 average properties, as determined from the analysis of the \ac{XMM} data. For the luminosities, the X-ray band (in keV) is indicated by the range in subscript.}
\label{table:abell_xray_luminosity}
\begin{tabular}{ll}
\toprule
    \multicolumn{2}{c}{Size} \\
    $R_\mathrm{500}$ & \multicolumn{1}{l}{(8.43\arcmin)} \\
    \midrule
    \multicolumn{2}{c}{Total mass} \\
    $M_\mathrm{500}$ & 6.664 (+0.401/-0.399) 10$^{13}$\Msun \\
    w. int. sc.    & 6.664 (+1.259/-1.543) 10$^{13}$\Msun \\
    \midrule
    \multicolumn{2}{c}{ICM quantities within $R_\mathrm{500}$} \\
    $T_\mathrm{spec}$           & 1.826 (+0.063/-0.065) keV \\
    $Z$             & 0.362 (+0.049/-0.044) $Z_\mathrm{\odot}$ \\
    $L_\mathrm{0.5-2.0}$ & \begin{tabular}[c]{@{}l@{}}
    1.913 (+0.036/-0.036) 10$^{36}$ W\end{tabular} \\
    $L_\mathrm{0.1-2.4}$ & \begin{tabular}[c]{@{}l@{}}
    3.165 (+0.073/-0.074) 10$^{36}$ W\end{tabular} \\
    $L_\mathrm{bol}$           & \begin{tabular}[c]{@{}l@{}}
    4.300 (+0.077/-0.077) 10$^{36}$ W\end{tabular} \\
\bottomrule
\end{tabular}
\end{table}

Because the cluster shows a central surface brightness peak, and seems to be in a relatively relaxed state, we calculated a projected temperature and abundance profile to see if there is a cool core. The X-ray temperature profile is shown in red and grey in Figure \ref{fig:xray_temp_profile}.

%%%%%%%%%
\subsection{Optical and infrared properties}
\label{subsection:Optical/IR properties}

\begin{table*}
\centering
    \caption{Properties of the Infrared and Optical Sources near the three Abell~S1136 Filaments.}
    \label{table:optical_and_ir_properties}
    \resizebox{\textwidth}{!}{%
    \begin{tabular}{clccccccccccl}
    \hline
    Source & \multicolumn{1}{c}{Source ID} & Flux Density & \multicolumn{8}{c}{Magnitude} & Redshift &  \\
    (Figure \ref{fig:1a_r0_contours}) & \multicolumn{1}{c}{} & \ac{ASKAP} (mJy){$^\dagger$} & \multicolumn{4}{c}{SkyMapper} & \multicolumn{4}{c}{WISEA} & ($z$){$^\star$} &  \\
     & \multicolumn{1}{c}{} &  & g & r & i & z & W1 & W2 & W3 & W1-W2 &  &  \\ \hline
     
    A & WISEA J233615.95$-$313534.4 & 1.24 & 16.197 & 15.617 & 15.135 & 14.845 & 12.396 & 12.445 & 11.699 & $-$0.049 & 0.0630 & \multicolumn{1}{c}{} \\
    \multicolumn{1}{l}{\medskip} & (SkyMapper J233615.95$-$313534.4) & $\pm$0.248 & $\pm$0.010 & $\pm$0.023 & $\pm$0.022 & $\pm$0.021 & $\pm$0.023 & $\pm$0.024 & $\pm$0.249 &  & $\pm$0.0021 &  \\
    
    B & WISEA J233616.55$-$313609.3 & 1.28 & 14.861 & 14.229 & 13.841 & 13.748 & 11.912 & 11.965 & 11.453 & $-$0.053 & 0.0622 & \multicolumn{1}{c}{} \\
    \multicolumn{1}{l}{\medskip} & (SkyMapper J233616.52$-$313609.0) & $\pm$0.256 & $\pm$0.128 & $\pm$0.091 & $\pm$0.087 & $\pm$0.133 & $\pm$0.023 & $\pm$0.023 & $\pm$0.187 & & $\pm$0.0031 &  \\
    
    C & WISEA J233616.10$-$313741.1	& 1.51 & $-$ & $-$ & $-$ & $-$ & 16.049 & 15.648 & 12.716 & 0.401 & $-$ & 
    \multicolumn{1}{c}{} \\
    \multicolumn{1}{l}{} &  & $\pm$0.302 &  &  &  &  & $\pm 0.061$ & $\pm 0.128$ & (upper limit) &  & \multicolumn{1}{l}{} &  \\ \hline
    \end{tabular}%
}
\begin{flushleft}
    {\scriptsize $^\dagger${~20\% uncertainty \citep{norris20}}. \\
    $^\star${~Redshift from 2dFGRS \citep{s1136_redshift_2dfs_2002MNRAS.329...87D}}.}
\end{flushleft}
\end{table*}

\begin{figure*}
\centering
    \includegraphics[trim=0 0 0 0, angle=270, width=0.8\textwidth]{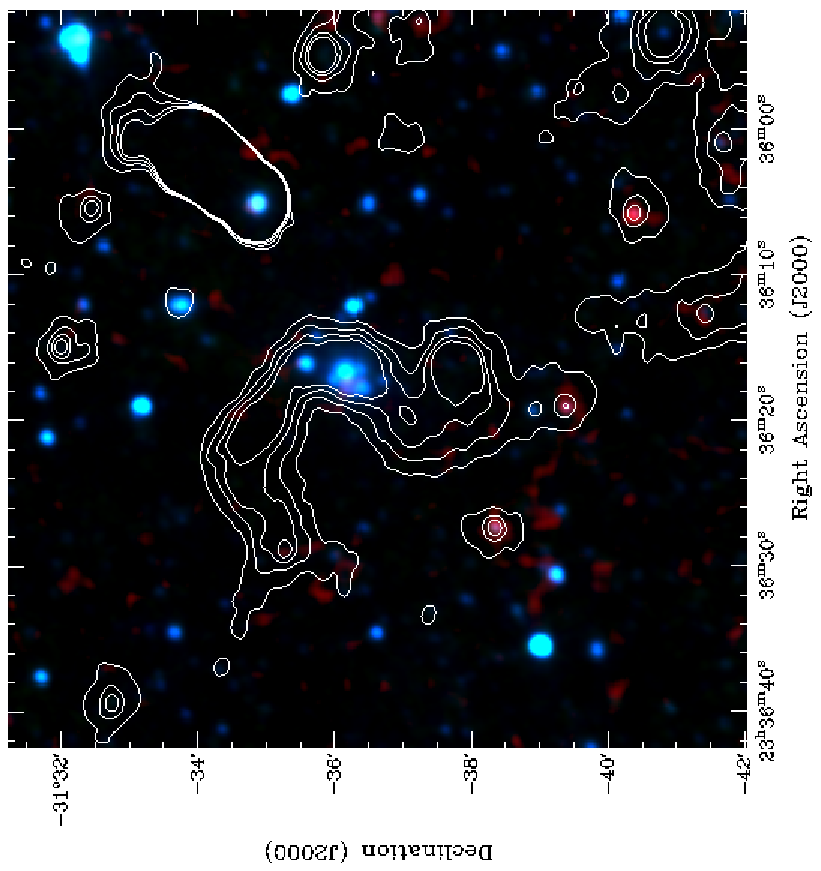}
    \caption[]{WISE infrared RGB image of the galaxy cluster Abell~S1136, overlaid with radio contours from the \ac{ASKAP} low resolution image (Figure \ref{fig:1b_r2_contours}; 25.51\arcsec$\times$21.27\arcsec). The RGB colours are red~=~22$\mu$m, green~=~4.6$\mu$m, blue~=~3.4$\mu$m. The \ac{BCG} ESO~470$-$G020 stands out near the image centre.}
    \label{fig:wise}
\end{figure*}

We show the radio contours overlaid on a WISE image in Figure \ref{fig:wise}, and in Table~\ref{table:optical_and_ir_properties} we list photometry and spectroscopy for sources A, B and C from the SkyMapper \citep{skymapper_dr2_2019PASA...36...33O} and WISE \citep{wise_2010AJ....140.1868W} surveys, and the 2dF Galaxy Redshift Survey \citep{s1136_redshift_2dfs_2002MNRAS.329...87D}. We classified the galaxies using  WISE colours following the WISE colour-colour diagram \citep{wise_2010AJ....140.1868W}.  

Source~A in Figure~\ref{fig:1a_r0_contours}, located near the south-west end of the  Northern Filament, is detected in both infrared (WISEA J233615.95$-$313534.4) and optical (SkyMapper J233615.95$-$313534.4). The WISE colours for this source (W1$-$W2 = $-$0.049, W2$-$W3 = 0.746) indicate this is most likely an elliptical galaxy. The redshift for this source is $z = 0.063 \pm0.002$, so it is a cluster member.

Source B in Figure \ref{fig:1a_r0_contours} is the \ac{BCG} of the Abell~S1136 galaxy cluster, and is coincident with the \ac{BCG} Filament.  This source is detected in both infrared (WISEA J233616.55$-$313609.3) and optical (SkyMapper J233616.52$-$313609.0). The WISE colours for this source (W1$-$W2 = $-$0.053, W2$-$W3 = 0.512) indicate this is most likely an elliptical galaxy. The redshift $z = 0.0622 \pm 0.0031$ for the Abell~S1136 galaxy cluster is measured on this source.

Source C in Figure \ref{fig:1a_r0_contours}, located slightly above the Southern Filament, is detected as a faint source in WISE bands W1 and W2 but not in other optical or IR photometry, giving little indication of the nature of this source, and no redshift is available.

%%%%%%%%%
\subsection{PKS 2333--318}
\label{subsection:pks2333-318}

\begin{figure*}
\centering
    \includegraphics[trim=0 0 0 0, width=0.8\textwidth]{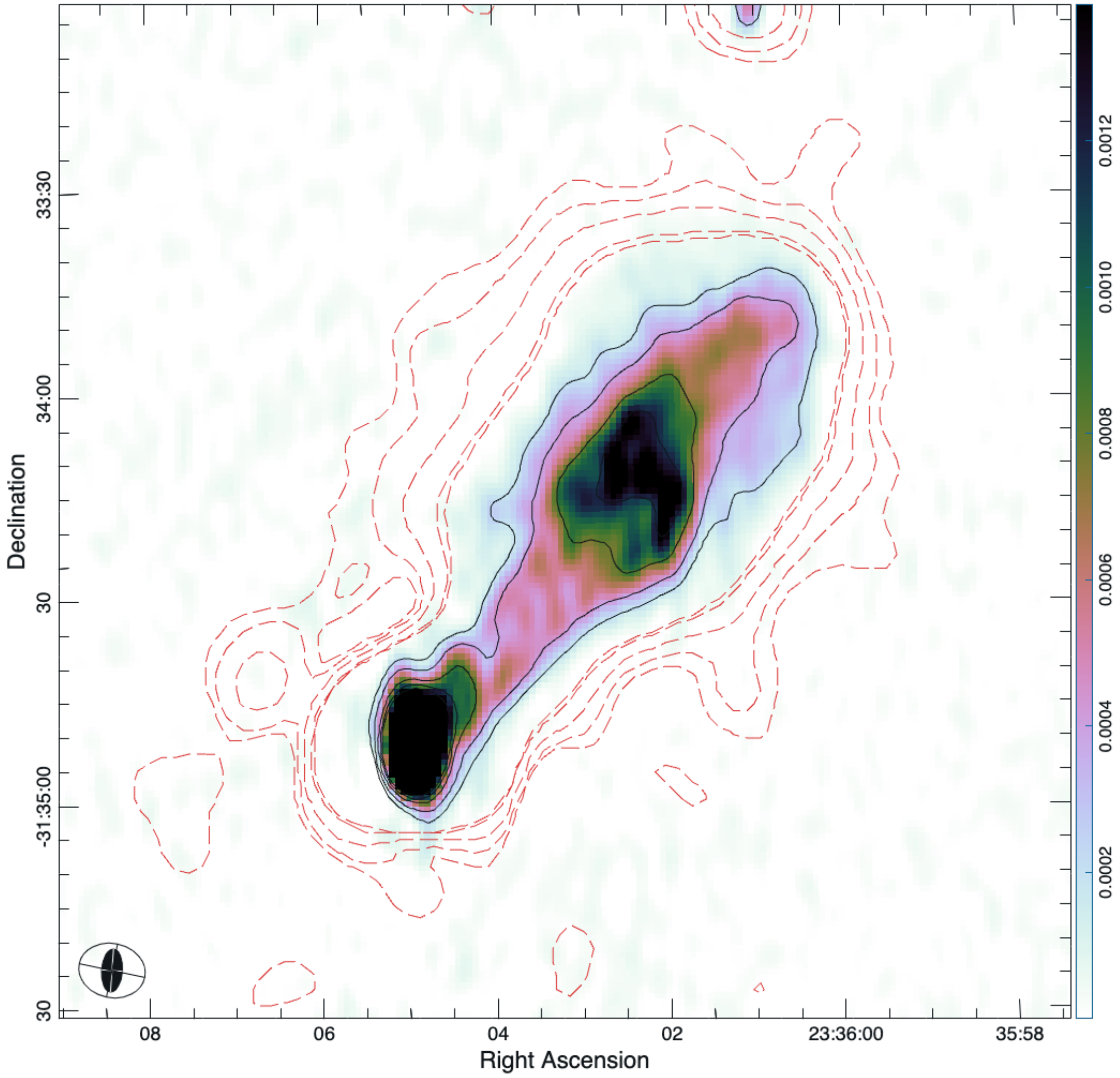}
    \caption[]{ATCA 2100~MHz radio continuum image (6.31\arcsec$\times$3.05\arcsec) of the interesting head-tail galaxy PKS~2333$-$318, located to the north-west of the Abell~S1136 galaxy cluster \ac{BCG}. The solid black contours at 200, 400, 800, 1200, and 1600~\ujybm, from the ATCA~2100~MHz image, outline the inner structure of the radio loud single tail emitted from PKS~2333$-$318.  The dashed red contours, at 80, 160, 320, 640, and 750~\ujybm, are from the \ac{ASKAP} {\tt{robust}\,$=-0.5$} image (12.6\arcsec$\times$10.0\arcsec) shown in Figure \ref{fig:1a_r0_contours}. The \ac{ATCA} and \ac{ASKAP} beams are shown in the bottom left corner as a solid and open ellipses, respectively.}
    \label{fig:pks2333-318}
\end{figure*}

The radio loud cluster member PKS~2333$-$318 lies to the north-west of the Abell~S1136 galaxy cluster at RA(J2000) = 23$^{\rm h}$36$^{\rm m}$04.96$^{\rm s}$ DEC(J2000) = $-$31\degr34\arcmin51.\arcsec3. The redshift of PKS~2333$-$318 is $z$ = 0.06134 \citep[][]{2009MNRAS.399..683J}. PKS~233$-$318 has an unusual single fat tail pointing away from the cluster centre. We consider this most likely to be a head-tail radio galaxy in which the jets have been blown back and merged by the \ac{ICM}. There is evidence of subtle sub-structure as shown in Figure \ref{fig:pks2333-318}, although there is no clear evidence of the double tail often seen in a head-tail galaxy.

%%%%%%%%% DISCUSSION %%%%%%%%%
\section{Discussion} 
\label{subsection:discussion}

\citet{Duchesne2021} detected diffuse emission from Abell~S1136 using the MWA. While they say the diffuse emission could be categorised as a cluster halo, potential alternative interpretations include the presence of a cluster relic obstructing the line of sight towards the cluster, or the possibility of a dormant radio galaxy that might have previously been associated with the \ac{BCG}, ESO~470$-$20. With the higher resolution and sensitivity of \ac{ASKAP} (Figure \ref{fig:1a_r0_contours}, we now see that the diffuse emission detected with the MWA consists of several components: (a) a cloud of diffuse emission about 450~kpc in extent; (b) three filaments of emission; and (c) a number of compact sources, one of which we identify as the \ac{BCG}. 

\subsection{The diffuse radio emission}

The diffuse emission (shown in green in Figure \ref{fig:detection}a) is about 450~kpc in extent. We consider whether the diffuse emission can be attributed to a radio halo, which is typically $\sim$~700~kpc to $\sim$~1~Mpc in size and found in disturbed clusters \citep[][]{chon_2012A&A...548A..59C, chon_2017A&A...606L...4C}, or to a mini-halo which is typically smaller; measuring less than 500 kpc in size, and found in relaxed clusters. 

Radio halos are probably caused by electrons accelerated by turbulence in the wake of cluster collisions. They are extended in size and trace the presence of cosmic rays and magnetic fields in the \ac{ICM},  and are seen in many merging galaxy clusters \citep[see e.g.][]{van-weeren-2019SSRv..215...16V}. However, both the mass and the X-ray luminosity of the Abell~S1136 galaxy cluster as reported here lie below any reported clusters with a radio halo (see eg Figure 9 in \citet[][]{van-weeren-2019SSRv..215...16V}). 

Radio mini-halos are normally found at the centre of relaxed cool-core clusters, and are associated with a radio galaxy that injects the electrons into the \ac{ICM}. Mini-halo sizes extend to $\sim$ 200 kpc from the central \ac{AGN}, with some evidence of radial spectral steepening as seen in the Perseus cluster \citep{gitti_2002A&A...386..456G}, Abell 2626 \citep{gitti_2004A&A...417....1G}, and Ophiucus \citep{murgia_2010A&A...514A..76M}. However, the higher sensitivity of the latest generation radio surveys is finding that many ``mini''-halos are far more extended than previously thought, with sizes up to approximately 0.5~Mpc; approaching the linear scales more commonly found for standard radio halos \citep[e.g.][]{savini-2018MNRAS.478.2234S,2019A&A...622A..24S,Biava2021-rxj1720,Riseley2022-ms1455,Riseley2023-a1413,Riseley2024-a2142}. 

In Section \ref{subsection:xray}, we showed that 
Abell~S1136 has some properties expected for relaxed clusters, but with no evidence of a cool core. The X-ray and radio emission were found to be coincident with the centre of the galaxy cluster, located on the \ac{BCG} ESO~470$-$20 \citep{bcg_eso-470-20_1989spce.book.....L}. 

The radio halo of a relaxed cluster classically consists of  a smooth volume-filling region of radio emission that roughly follows the mass distribution of the cluster, and the radio halo centroid is coincident with the centroid of the X-ray emission, usually with some form of radial symmetry. The diffuse radio emission in Abell~S1136, while roughly centred on the \ac{BCG} and on the centroid of the X-ray emission, is far from being a contiguous or symmetric smooth region about the \ac{BCG}. 

The measured spectral index ($\alpha\sim-$1.68) of the diffuse emission in Abell~S1136 is typical for radio halos or their smaller mini-halo counterparts ($\alpha<-$1). Revived fossil plasma sources often show steep spectral indices ($\alpha\lesssim-2$), and a curved spectrum \citep{Riseley2022_Abell3266}; we see in Figure \ref{fig_spix_curvature} the MWA~2 data at low frequencies exhibits a possible break in the simple power law fit, with the spectral index being steeper at the lower frequencies.
    
Although there might be a curved spectrum in the diffuse emission of the Abell~S1136 galaxy cluster, as seen in revived fossil plasma, we are unable to quantify this as there are no measurements between the \ac{MWA} 215~MHz and \ac{ASKAP} 888~MHz data points. Therefore, the properties of the diffuse emission do not fit well with either classifications of a giant, or mini, radio halo.

We therefore cannot classify the source as a halo or mini-halo, and defer its classification to the wider discussion  of the classification  of  diffuse emission from clusters \citep[][]{2019Sci...364..981G, 2020MNRAS.499L..11B, 2022A&A...657A..56K}, prompted by  higher resolution and sensitivity observations using telescopes such as \ac{ASKAP}, MeerKAT, and \ac{LOFAR}.

\subsection{The X-ray channel}
\label{channel}

Figure~\ref{fig:askap_and_xmm} shows marginal evidence of a ``channel'' through the hot gas of the \ac{ICM}, coincident with the Northern filament. To calculate the statistical significance of this channel, we define a control region (shown in Figure~\ref{fig:askap_and_xmm}) consisting of an annulus between 1 and 2\arcmin\ around the cluster centre, excluding a region immediately surrounding the potential channel. We then fit a radial power law to the flux distribution inside the control annulus, corrected for various small effects, on which we simulate the channel using a  MCMC (Markov chain Monte Carlo) with 50 million points. This resulted in an estimated probability of obtaining the observed channel by chance that corresponds to $\sim 4.3 \sigma$, which we regard as ``marginally significant''.
   
We therefore acknowledge that this channel may  just be a statistical fluctuation in the noise.  For example, other similar ``holes'' in the X-ray surface brightness can be seen in Figure~\ref{fig:askap_and_xmm}, possibly due to the intrinsic variance of the underlying ICM distribution. Furthermore, there is no evidence for an overall anti-correlation between radio and X-ray. However, the alignment with the radio is curious, and therefore merits some discussion. If this channel is real, it might be caused by the interaction of the  radio jet from an AGN  with the hot X-ray gas in the \ac{ICM}. Examples of this have been seen in NGC 1275 \citep{radio_jet_punching_ngc_1275_1993MNRAS.264L..25B}, \citep{A194} and possibly in Abell~S0102 \citep{S0102_2001A&A...369..467R, 2010SerAJ.181...31F}. However, it is unclear whether the radio emission in the current case is confined by the X-ray medium, with the X-ray gas guiding the radio emission, or is the result of a channel being bored by the radio jet \citep{radio_channel_2017PhPl...24d1402J} through the hot gas. In either case, the filament would then presumably represent a  radio jet from source~A. However, we are cautious about over-interpreting this channel as it is of marginal statistical significance (4.3 $\sigma$). 
 
%%%%%%%%% 
\subsection{Filaments}

Filamentary radio emission in clusters is typically caused by radio shocks, as discussed in Section~\ref{section:introduction}, and are typically aligned tangentially around the periphery of the cluster \citep{van-weeren-2019SSRv..215...16V}; With the possible exception of the Southern Filament, the three filaments that we have identified in the Abell~S1136 do not follow this pattern. In some cases the filamentary emission can be at the periphery of the cluster, but seen in projection - eg along the line of sight through the cluster, so the periphery is facing us; this might be the case for the Southern Filament. 

The Northern Filament is $\sim$140~kpc long and is oriented towards the radio galaxy~A, which is at a similar redshift to the \ac{BCG} galaxy~B.  As discussed above in Section~\ref{channel}, there is  marginal evidence for an X-ray channel coincident with this filament, suggesting the filament may represent a fossil radio jet from source~A. Alternatively, since the filament is near the periphery of the cluster, it could be a radio shock.

The \ac{BCG} filament appears to extend from the \ac{BCG}, and is $\sim$140~kpc long. This is most likely to be a fossil jet associated with galaxy~B, although we cannot discount the possibility that it might be a section of an unusual \ac{WAT}.

The Southern Filament is only $\sim$80~kpc long, and may be associated with with radio source C, in which case it might be revived fossil plasma. However, the filament is perpendicular to the radius vector from the \ac{BCG}, suggesting that it may be a radio shock. 
No redshift is available for the radio source C, and therefore we do not know whether or not this source is associated with the cluster. 

With new deep and high resolution low-frequency observations, such as \ac{uGMRT} and MeerKAT, the existence, morphology, and spectral nature of these filaments could be confirmed and studied in more detail. Additionally, deep linear polarisation data such as from the \ac{ASKAP} Polarisation Sky Survey of the Universe's Magnetism \citep[POSSUM;][]{2010AAS...21547013G} could reveal the polarisation fraction and orientation of the magnetic fields which may help discern whether these are shocks are not.

%%%%%%%%% 
\subsection{X-ray}

In our analysis of the X-ray temperatures in the Abell~S1136 cluster, we do not see cooler emission in the central parts, which would be characteristic of a cool core cluster \citep{cool_core_temp_drop_2010A&A...513A..37H}. Instead the central parts appear significantly hotter, which would indicate either that the cluster is not relaxed, or that the central AGN inside the \ac{BCG} also emits X-rays, which would harden the spectra. However,  we found no evidence for \ac{AGN} emission. The reduced $\chi^2$ for the spectral modelling of the central bin is not changed significantly by adding a power-law component to account for a central AGN, which does not allow us to dismiss or confirm this hypothesis. However, we note that the best fit AGN contribution would be 10\% of the central flux and would flatten the central profile to temperature values comparable to the measurements at 0.2$-$0.3 $R_\mathrm{500}$, which is more in line with expectations for a cluster without evidence for a strong merger event. It would also explain quite well the central surface brightness peak in a cluster without a cool core.

%%%%%%%%% CONCLUSION %%%%%%%%%
\section{Conclusions}
\label{section:conclusion}
    
The diffuse radio emission in the Abell~S1136 galaxy cluster appeared in earlier observations to be an amorphous radio blob, similar in appearance to a radio halo. 

Our higher-resolution and higher-sensitivity \ac{ASKAP} observations show that it breaks up into: (a) a region of diffuse emission about 450~kpc in extent; (b) three filaments located within the diffuse emission, each 80$-$140~kpc in extent; and (c) a small number of compact radio sources. The diffuse emission shows some structure which was not evident in earlier low-resolution observations, and its properties do not closely match either those of a halo or mini-halo. However, the distinction between halos and mini-halos is becoming less clear in other high-sensitivity observations of clusters. 

The three filaments appear similar to those increasingly being found in high-sensitivity observations of other cluster sources, although we cannot rule out the possibility that they represent an unusual \ac{WAT}. Further investigation using higher quality spectral index data and polarisation information would be helpful.

Our X-ray analysis of Abell S1136, based on XMM-Newton observations, shows a distinctive X-ray emission pattern closely aligned with the radio emission from the \ac{BCG}. This alignment reinforces the BCG's role as both the geometric centre and the locus of the cluster's X-ray emission. There is a possible ``channel'' in the X-ray plasm, which aligns with the ``Northern'' filament; however evidence for the channel is only marginally significant at $4.3\sigma$. While it is interesting, it may just signify a statistical fluctuation. The X-ray temperature profile shows that the cluster appears to be in a relaxed state, with no evidence of a cool core. 

We note that these filaments, and the structure in the diffuse emission, were not visible in earlier, low-resolution observations, and we speculate that many cluster radio sources which are currently regarded as smooth extended sources may well show more complex structure, such as that we see in Abell~S1136, when observed with next-generation radio telescopes such as \ac{ASKAP}, MeerKAT, \ac{uGMRT} and \ac{LOFAR}. The wealth of such details that we can expect in the next few years may prompt a re-examination of the classifications of the radio properties of clusters that have been based on low-resolution data.

%%%%%%%%% ACKNOWLEDGEMENTS %%%%%%%%%
\section*{Acknowledgements}

This scientific work uses data obtained from Ilyarrimanha Ilgari Bundara, the CSIRO Murchison Radio-astronomy Observatory. We acknowledge the Wajarri Yamaji People as the Traditional Owners and native title holders of the Observatory site. The Australian SKA Pathfinder is part of the Australia Telescope National Facility (\url{https://ror.org/05qajvd42}) which is managed by CSIRO. Operation of ASKAP is funded by the Australian Government with support from the National Collaborative Research Infrastructure Strategy. ASKAP uses the resources of the Pawsey Supercomputing Centre. Establishment of ASKAP, the Murchison Radio-astronomy Observatory and the Pawsey Supercomputing Centre are initiatives of the Australian Government, with support from the Government of Western Australia and the Science and Industry Endowment Fund.

The Australia Telescope Compact Array is part of the Australia Telescope National Facility (\url{https://ror.org/05qajvd42}) which is funded by the Australian Government for operation as a National Facility managed by CSIRO. We acknowledge the Gomeroi people as the traditional owners of the Observatory site. 

This paper includes archived data obtained through the Australia Telescope Online Archive (\url{http://atoa.atnf.csiro.au}). 

Partial support for LR comes from U.S. National Science Foundation grant AST17-14205 to the University of Minnesota. SWD acknowledges an Australian Government Research Training Program scholarship administered through Curtin University. CJR acknowledges financial support from the ERC Starting Grant ``DRANOEL'', number 714245. MA acknowledges the financial support from the European Union - NextGenerationEU and the Spanish Ministry of Science and Innovation through the Recovery and Resilience Facility project J-CAVA, as well as The State Research Agency (AEI-MCINN) of the Spanish Ministry of Science and Innovation (SMSI) under the grant \\ 
PID2019-105602GBI00/10.13039/501100011033 and the IAC Project P/300724, financed by the SMSI, through the Canary Islands Department of Economy, Knowledge and Employment.

This work was partly done using GNU Astronomy Utilities (Gnuastro, ascl.net/1801.009) version \maGnuastroVersion. Work on Gnuastro has been funded by the Japanese Ministry of Education, Culture, Sports, Science, and Technology (MEXT) scholarship and its Grant-in-Aid for Scientific Research (21244012, 24253003), the European Research Council (ERC) advanced grant 339659-MUSICOS, and from the Spanish Ministry of Economy and Competitiveness (MINECO) under grant number AYA2016-76219-P. 

Additional data processing and analysis were conducted using the \textsc{karma}\footnote{\url{https://www.atnf.csiro.au/computing/software/karma/}} \citep{kvis_karma_1995ASPC...77..144G} software visualisation package. 

%%%%%%%%% DATA AVAILABILITY %%%%%%%%%
\section*{Data Availability}

Stokes I  Taylor term images of the \ac{ASKAP} data used in this article are available through the CSIRO \ac{ASKAP} Science Data Archive (CASDA) under \url{https://doi.org/10.25919/44sn-2x47}{}. The \ac{ATCA} data are available from the Australia Telescope Online Archive at \url{https://atoa.atnf.csiro.au/query.jsp}. DSS and WISE images were obtained through the SkyView\footnote{\url{https://skyview.gsfc.nasa.gov/current/cgi/query.pl}} servers. The X-ray data were obtained from the XMM-Newton Science Archive (XSA)\footnote{\url{https://www.cosmos.esa.int/web/xmm-newton/xsa}}. The MWA~I data are available from the MWA node of the All Sky Virtual Observatory (ASVO)\footnote{\url{https://wiki.mwatelescope.org/display/MP/Data+Access}}. The MWA~2 data are available from the corresponding authors upon reasonable request. Catalogue information was obtained from SkyMapper\footnote{\url{http://skymapper.anu.edu.au/}} and VizieR\footnote{\url{https://vizier.u-strasbg.fr/viz-bin/VizieR}}.

The code to reproduce the NoiseChisel results in Section \ref{subsection_noisechisel} is available at \url{https://gitlab.com/makhlaghi/abell-s1136}.

%%%%%%%%% REFERENCES %%%%%%%%%
\bibliography{s1136}

\begin{thebibliography}{}
\expandafter\ifx\csname natexlab\endcsname\relax\def\natexlab#1{#1}\fi

\bibitem[{{Akhlaghi}(2019{\natexlab{a}})}]{akhlaghi21}
{Akhlaghi}, M. 2019{\natexlab{a}}, arXiv e-prints, arXiv:1909.11230

\bibitem[{{Akhlaghi}(2019{\natexlab{b}})}]{akhlaghi19}
{Akhlaghi}, M. 2019{\natexlab{b}}, in Astronomical Society of the Pacific Conference Series, Vol. 521, Astronomical Data Analysis Software and Systems XXVI, ed. M.~{Molinaro}, K.~{Shortridge}, \& F.~{Pasian}, 299

\bibitem[{{Akhlaghi} \& {Ichikawa}(2015)}]{akhlaghi15}
{Akhlaghi}, M., \& {Ichikawa}, T. 2015, ApJS, 220, 1

\bibitem[{{Arnaud} {et~al.}(2005){Arnaud}, {Pointecouteau}, \& {Pratt}}]{arnaud_2005A&A...441..893A}
{Arnaud}, M., {Pointecouteau}, E., \& {Pratt}, G.~W. 2005, \aap, 441, 893

\bibitem[{{Beardsley} {et~al.}(2019){Beardsley}, {Johnston-Hollitt}, {Trott}, {Pober}, {Morgan}, {Oberoi}, {Kaplan}, {Lynch}, {Anderson}, {McCauley}, {Croft}, {James}, {Wong}, {Tremblay}, {Norris}, {Cairns}, {Lonsdale}, {Hancock}, {Gaensler}, {Bhat}, {Li}, {Hurley-Walker}, {Callingham}, {Seymour}, {Yoshiura}, {Joseph}, {Takahashi}, {Sokolowski}, {Miller-Jones}, {Chauhan}, {Boji{\v{c}}i{\'c}}, {Filipovi{\'c}}, {Leahy}, {Su}, {Tian}, {McSweeney}, {Meyers}, {Kitaeff}, {Vernstrom}, {G{\"u}rkan}, {Heald}, {Xue}, {Riseley}, {Duchesne}, {Bowman}, {Jacobs}, {Crosse}, {Emrich}, {Franzen}, {Horsley}, {Kenney}, {Morales}, {Pallot}, {Steele}, {Tingay}, {Walker}, {Wayth}, {Williams}, \& {Wu}}]{2019PASA...36...50B}
{Beardsley}, A.~P., {Johnston-Hollitt}, M., {Trott}, C.~M., {et~al.} 2019, \pasa, 36, e050

\bibitem[{{Biava} {et~al.}(2021){Biava}, {de Gasperin}, {Bonafede}, {Edler}, {Giacintucci}, {Mazzotta}, {Brunetti}, {Botteon}, {Br{\"u}ggen}, {Cassano}, {Drabent}, {Edge}, {En{\ss}lin}, {Gastaldello}, {Riseley}, {Rossetti}, {Rottgering}, {Shimwell}, {Tasse}, \& {van Weeren}}]{Biava2021-rxj1720}
{Biava}, N., {de Gasperin}, F., {Bonafede}, A., {et~al.} 2021, \mnras, 508, 3995

\bibitem[{{Biava} {et~al.}(2024){Biava}, {Bonafede}, {Gastaldello}, {Botteon}, {Brienza}, {Shimwell}, {Brunetti}, {Bruno}, {Rajpurohit}, {Riseley}, {van Weeren}, {Rossetti}, {Cassano}, {De Gasperin}, {Drabent}, {Rottgering}, {Edge}, \& {Tasse}}]{Biava2024-cool-core-sample}
{Biava}, N., {Bonafede}, A., {Gastaldello}, F., {et~al.} 2024, arXiv e-prints, arXiv:2403.09802

\bibitem[{{Bock} {et~al.}(1999){Bock}, {Large}, \& {Sadler}}]{sumss-bock-1999AJ....117.1578B}
{Bock}, D.~C.~J., {Large}, M.~I., \& {Sadler}, E.~M. 1999, \aj, 117, 1578

\bibitem[{{Boehringer} {et~al.}(1993){Boehringer}, {Voges}, {Fabian}, {Edge}, \& {Neumann}}]{radio_jet_punching_ngc_1275_1993MNRAS.264L..25B}
{Boehringer}, H., {Voges}, W., {Fabian}, A.~C., {Edge}, A.~C., \& {Neumann}, D.~M. 1993, \mnras, 264, L25

\bibitem[{{Bonafede} {et~al.}(2010){Bonafede}, {Feretti}, {Murgia}, {Govoni}, {Giovannini}, {Dallacasa}, {Dolag}, \& {Taylor}}]{2010A&A...513A..30B}
{Bonafede}, A., {Feretti}, L., {Murgia}, M., {et~al.} 2010, \aap, 513, A30

\bibitem[{{Bonafede} {et~al.}(2014){Bonafede}, {Intema}, {Bruggen}, {Russell}, {Ogrean}, {Basu}, {Sommer}, {van Weeren}, {Cassano}, {Fabian}, \& {Rottgering}}]{2014MNRAS.444L..44B}
{Bonafede}, A., {Intema}, H.~T., {Bruggen}, M., {et~al.} 2014, \mnras, 444, L44

\bibitem[{{Botteon} {et~al.}(2020){Botteon}, {van Weeren}, {Brunetti}, {de Gasperin}, {Intema}, {Osinga}, {Di Gennaro}, {Shimwell}, {Bonafede}, {Br{\"u}ggen}, {Cassano}, {Cuciti}, {Dallacasa}, {Gastaldello}, {Mandal}, {Rossetti}, \& {R{\"o}ttgering}}]{2020MNRAS.499L..11B}
{Botteon}, A., {van Weeren}, R.~J., {Brunetti}, G., {et~al.} 2020, \mnras, 499, L11

\bibitem[{{Botteon} {et~al.}(2022){Botteon}, {Shimwell}, {Cassano}, {Cuciti}, {Zhang}, {Bruno}, {Camillini}, {Natale}, {Jones}, {Gastaldello}, {Simionescu}, {Rossetti}, {Akamatsu}, {van Weeren}, {Brunetti}, {Br{\"u}ggen}, {Groeneveld}, {Hoang}, {Hardcastle}, {Ignesti}, {Di Gennaro}, {Bonafede}, {Drabent}, {R{\"o}ttgering}, {Hoeft}, \& {de Gasperin}}]{botteon_halos_2022A&A...660A..78B}
{Botteon}, A., {Shimwell}, T.~W., {Cassano}, R., {et~al.} 2022, \aap, 660, A78

\bibitem[{{Bowman} {et~al.}(2013){Bowman}, {Cairns}, {Kaplan}, {Murphy}, {Oberoi}, {Staveley-Smith}, {Arcus}, {Barnes}, {Bernardi}, {Briggs}, {Brown}, {Bunton}, {Burgasser}, {Cappallo}, {Chatterjee}, {Corey}, {Coster}, {Deshpande}, {deSouza}, {Emrich}, {Erickson}, {Goeke}, {Gaensler}, {Greenhill}, {Harvey-Smith}, {Hazelton}, {Herne}, {Hewitt}, {Johnston-Hollitt}, {Kasper}, {Kincaid}, {Koenig}, {Kratzenberg}, {Lonsdale}, {Lynch}, {Matthews}, {McWhirter}, {Mitchell}, {Morales}, {Morgan}, {Ord}, {Pathikulangara}, {Prabu}, {Remillard}, {Robishaw}, {Rogers}, {Roshi}, {Salah}, {Sault}, {Shankar}, {Srivani}, {Stevens}, {Subrahmanyan}, {Tingay}, {Wayth}, {Waterson}, {Webster}, {Whitney}, {Williams}, {Williams}, \& {Wyithe}}]{bowman_2013PASA...30...31B}
{Bowman}, J.~D., {Cairns}, I., {Kaplan}, D.~L., {et~al.} 2013, \pasa, 30, e031

\bibitem[{{Brentjens}(2008)}]{2008A&A...489...69B}
{Brentjens}, M.~A. 2008, \aap, 489, 69

\bibitem[{{Briggs}(1995)}]{Briggs1995}
{Briggs}, D.~S. 1995, PhD thesis, The New Mexico Institute of Mining and Technology

\bibitem[{{Brunetti} {et~al.}(2009){Brunetti}, {Cassano}, {Dolag}, \& {Setti}}]{brunetti_giant_halo_evolution_2009A&A...507..661B}
{Brunetti}, G., {Cassano}, R., {Dolag}, K., \& {Setti}, G. 2009, \aap, 507, 661

\bibitem[{{Campusano} {et~al.}(2018){Campusano}, {Marinello}, {Clowes}, {Haines}, {Pereira}, {Pizarro}, {Hitschfeld-Kahler}, \& {S{\"o}chting}}]{redshift_2018ApJ...869..145C}
{Campusano}, L.~E., {Marinello}, G., {Clowes}, R.~G., {et~al.} 2018, \apj, 869, 145

\bibitem[{{Carilli} \& {Taylor}(2002)}]{2002ARA&A..40..319C}
{Carilli}, C.~L., \& {Taylor}, G.~B. 2002, \araa, 40, 319

\bibitem[{{Cassano} {et~al.}(2008){Cassano}, {Gitti}, \& {Brunetti}}]{cassano_mini_halo_emission_2008A&A...486L..31C}
{Cassano}, R., {Gitti}, M., \& {Brunetti}, G. 2008, \aap, 486, L31

\bibitem[{{Chon} \& {B{\"o}hringer}(2017)}]{chon_2017A&A...606L...4C}
{Chon}, G., \& {B{\"o}hringer}, H. 2017, \aap, 606, L4

\bibitem[{{Chon} {et~al.}(2012){Chon}, {B{\"o}hringer}, \& {Smith}}]{chon_2012A&A...548A..59C}
{Chon}, G., {B{\"o}hringer}, H., \& {Smith}, G.~P. 2012, \aap, 548, A59

\bibitem[{{Clarke} {et~al.}(2001){Clarke}, {Kronberg}, \& {B{\"o}hringer}}]{2001ApJ...547L.111C}
{Clarke}, T.~E., {Kronberg}, P.~P., \& {B{\"o}hringer}, H. 2001, \apjl, 547, L111

\bibitem[{{Comrie} {et~al.}(2021){Comrie}, {Wang}, {Hsu}, {Moraghan}, {Harris}, {Pang}, {Pi{\'n}ska}, {Chiang}, {Chang}, {Hwang}, {Jan}, {Lin}, \& {Simmonds}}]{carta}
{Comrie}, A., {Wang}, K.-S., {Hsu}, S.-C., {et~al.} 2021, {CARTA: The Cube Analysis and Rendering Tool for Astronomy}, doi:\url{10.5281/zenodo.3377984}

\bibitem[{{Condon} {et~al.}(1998){Condon}, {Cotton}, {Greisen}, {Yin}, {Perley}, {Taylor}, \& {Broderick}}]{nvss-1998AJ....115.1693C}
{Condon}, J.~J., {Cotton}, W.~D., {Greisen}, E.~W., {et~al.} 1998, \aj, 115, 1693

\bibitem[{{Cuciti} {et~al.}(2021{\natexlab{a}}){Cuciti}, {Cassano}, {Brunetti}, {Dallacasa}, {van Weeren}, {Giacintucci}, {Bonafede}, {de Gasperin}, {Ettori}, {Kale}, {Pratt}, \& {Venturi}}]{cuciti_halos_and_mergers_1_2021A&A...647A..50C}
{Cuciti}, V., {Cassano}, R., {Brunetti}, G., {et~al.} 2021{\natexlab{a}}, \aap, 647, A50

\bibitem[{{Cuciti} {et~al.}(2021{\natexlab{b}}){Cuciti}, {Cassano}, {Brunetti}, {Dallacasa}, {de Gasperin}, {Ettori}, {Giacintucci}, {Kale}, {Pratt}, {van Weeren}, \& {Venturi}}]{cuciti_halos_and_mergers_2_2021A&A...647A..51C}
---. 2021{\natexlab{b}}, \aap, 647, A51

\bibitem[{{De Propris} {et~al.}(2002){De Propris}, {Couch}, {Colless}, {Dalton}, {Collins}, {Baugh}, {Bland -Hawthorn}, {Bridges}, {Cannon}, {Cole}, {Cross}, {Deeley}, {Driver}, {Efstathiou}, {Ellis}, {Frenk}, {Glazebrook}, {Jackson}, {Lahav}, {Lewis}, {Lumsden}, {Maddox}, {Madgwick}, {Moody}, {Norberg}, {Peacock}, {Percival}, {Peterson}, {Sutherland}, \& {Taylor}}]{s1136_redshift_2dfs_2002MNRAS.329...87D}
{De Propris}, R., {Couch}, W.~J., {Colless}, M., {et~al.} 2002, \mnras, 329, 87

\bibitem[{{Donnert} {et~al.}(2018){Donnert}, {Vazza}, {Br{\"u}ggen}, \& {ZuHone}}]{donnert_2018SSRv..214..122D}
{Donnert}, J., {Vazza}, F., {Br{\"u}ggen}, M., \& {ZuHone}, J. 2018, \ssr, 214, 122

\bibitem[{{Duchesne} {et~al.}(2021){Duchesne}, {Johnston-Hollitt}, {Offringa}, {Pratt}, {Zheng}, \& {Dehghan}}]{Duchesne2021}
{Duchesne}, S.~W., {Johnston-Hollitt}, M., {Offringa}, A.~R., {et~al.} 2021, \pasa, 38, e010

\bibitem[{{Duchesne} {et~al.}(2020){Duchesne}, {Johnston-Hollitt}, {Zhu}, {Wayth}, \& {Line}}]{Duchesne2020}
{Duchesne}, S.~W., {Johnston-Hollitt}, M., {Zhu}, Z., {Wayth}, R.~B., \& {Line}, J.~L.~B. 2020, \pasa, 37, e037

\bibitem[{{Ensslin} {et~al.}(1998){Ensslin}, {Biermann}, {Klein}, \& {Kohle}}]{1998A&A...332..395E}
{Ensslin}, T.~A., {Biermann}, P.~L., {Klein}, U., \& {Kohle}, S. 1998, \aap, 332, 395

\bibitem[{{Feretti} {et~al.}(2012){Feretti}, {Giovannini}, {Govoni}, \& {Murgia}}]{taxonomy_feretti_2012A&ARv..20...54F}
{Feretti}, L., {Giovannini}, G., {Govoni}, F., \& {Murgia}, M. 2012, \aapr, 20, 54

\bibitem[{{Filipovic} {et~al.}(2010){Filipovic}, {Crawford}, {Jones}, \& {White}}]{2010SerAJ.181...31F}
{Filipovic}, M.~D., {Crawford}, E.~J., {Jones}, P.~A., \& {White}, G.~L. 2010, Serbian Astronomical Journal, 181, 31

\bibitem[{{Filipovi{\'c}} \& {Tothill}(2021)}]{2021pma..book.....F}
{Filipovi{\'c}}, M.~D., \& {Tothill}, N. F.~H. 2021, {Principles of Multimessenger Astronomy} ({IOP Publishing}), doi:\url{10.1088/2514-3433/ac087e}

\bibitem[{{Frater} {et~al.}(1992){Frater}, {Brooks}, \& {Whiteoak}}]{atca-frater}
{Frater}, R.~H., {Brooks}, J.~W., \& {Whiteoak}, J.~B. 1992, Journal of Electrical and Electronics Engineering Australia, 12, 103

\bibitem[{{Gaensler} {et~al.}(2010){Gaensler}, {Landecker}, {Taylor}, \& {POSSUM Collaboration}}]{2010AAS...21547013G}
{Gaensler}, B.~M., {Landecker}, T.~L., {Taylor}, A.~R., \& {POSSUM Collaboration}. 2010, in American Astronomical Society Meeting Abstracts, Vol. 215, American Astronomical Society Meeting Abstracts \#215, 470.13

\bibitem[{{Ghirardini} {et~al.}(2019){Ghirardini}, {Eckert}, {Ettori}, {Pointecouteau}, {Molendi}, {Gaspari}, {Rossetti}, {De Grandi}, {Roncarelli}, {Bourdin}, {Mazzotta}, {Rasia}, \& {Vazza}}]{2019A&A...621A..41G}
{Ghirardini}, V., {Eckert}, D., {Ettori}, S., {et~al.} 2019, \aap, 621, A41

\bibitem[{{Giacintucci} {et~al.}(2008){Giacintucci}, {Venturi}, {Macario}, {Dallacasa}, {Brunetti}, {Markevitch}, {Cassano}, {Bardelli}, \& {Athreya}}]{Giacintucci-2008A&A...486..347G}
{Giacintucci}, S., {Venturi}, T., {Macario}, G., {et~al.} 2008, \aap, 486, 347

\bibitem[{{Giovannini} \& {Feretti}(2004)}]{2004JKAS...37..323G}
{Giovannini}, G., \& {Feretti}, L. 2004, Journal of Korean Astronomical Society, 37, 323

\bibitem[{{Giovannini} {et~al.}(1999){Giovannini}, {Tordi}, \& {Feretti}}]{giovanniniRadiHaloAndRelicCandidatesNRAO-1999NewA....4..141G}
{Giovannini}, G., {Tordi}, M., \& {Feretti}, L. 1999, \na, 4, 141

\bibitem[{{Gitti} {et~al.}(2004){Gitti}, {Brunetti}, {Feretti}, \& {Setti}}]{gitti_2004A&A...417....1G}
{Gitti}, M., {Brunetti}, G., {Feretti}, L., \& {Setti}, G. 2004, \aap, 417, 1

\bibitem[{{Gitti} {et~al.}(2002){Gitti}, {Brunetti}, \& {Setti}}]{gitti_2002A&A...386..456G}
{Gitti}, M., {Brunetti}, G., \& {Setti}, G. 2002, \aap, 386, 456

\bibitem[{{Gitti} {et~al.}(2015){Gitti}, {Tozzi}, {Brunetti}, {Cassano}, {Dallacasa}, {Edge}, {Ettori}, {Feretti}, {Ferrari}, {Giacintucci}, {Giovannini}, {Hogan}, \& {Venturi}}]{gitti_mini_halos_2015aska.confE..76G}
{Gitti}, M., {Tozzi}, P., {Brunetti}, G., {et~al.} 2015, in Advancing Astrophysics with the Square Kilometre Array (AASKA14), 76

\bibitem[{{Gooch}(1995)}]{kvis_karma_1995ASPC...77..144G}
{Gooch}, R. 1995, Astronomical Society of the Pacific Conference Series, Vol.~77, {Space and the Spaceball} (Astronomical Society of the Pacific), 144

\bibitem[{{G{\'o}rski} {et~al.}(2005){G{\'o}rski}, {Hivon}, {Banday}, {Wandelt}, {Hansen}, {Reinecke}, \& {Bartelmann}}]{healpix_2005ApJ...622..759G}
{G{\'o}rski}, K.~M., {Hivon}, E., {Banday}, A.~J., {et~al.} 2005, \apj, 622, 759

\bibitem[{{Govoni} {et~al.}(2019){Govoni}, {Orr{\`u}}, {Bonafede}, {Iacobelli}, {Paladino}, {Vazza}, {Murgia}, {Vacca}, {Giovannini}, {Feretti}, {Loi}, {Bernardi}, {Ferrari}, {Pizzo}, {Gheller}, {Manti}, {Br{\"u}ggen}, {Brunetti}, {Cassano}, {de Gasperin}, {En{\ss}lin}, {Hoeft}, {Horellou}, {Junklewitz}, {R{\"o}ttgering}, {Scaife}, {Shimwell}, {van Weeren}, \& {Wise}}]{2019Sci...364..981G}
{Govoni}, F., {Orr{\`u}}, E., {Bonafede}, A., {et~al.} 2019, Science, 364, 981

\bibitem[{{Guzman} {et~al.}(2019){Guzman}, {Whiting}, {Voronkov}, {Mitchell}, {Ord}, {Collins}, {Marquarding}, {Lahur}, {Maher}, {Van Diepen}, {Bannister}, {Wu}, {Lenc}, {Khoo}, \& {Bastholm}}]{askapsoft_2019ascl.soft12003G}
{Guzman}, J., {Whiting}, M., {Voronkov}, M., {et~al.} 2019, {ASKAPsoft: ASKAP science data processor software}, ascl:1912.003

\bibitem[{Hindson {et~al.}(2014)Hindson, Johnston-Hollitt, Hurley-Walker, Buckley, Morgan, Carretti, Dwarakanath, Bell, Bernardi, Bhat, Bowman, Briggs, Cappallo, Corey, Deshpande, Emrich, Ewall-Wice, Feng, Gaensler, Goeke, Greenhill, Hazelton, Jacobs, Kaplan, Kasper, Kratzenberg, Kudryavtseva, Lenc, Lonsdale, Lynch, McWhirter, McKinley, Mitchell, Morales, Morgan, Oberoi, Ord, Pindor, Prabu, Procopio, Offringa, Riding, Rogers, Roshi, Shankar, Srivani, Subrahmanyan, Tingay, Waterson, Wayth, Webster, Whitney, Williams, \& Williams}]{Hindson-10.1093/mnras/stu1669}
Hindson, L., Johnston-Hollitt, M., Hurley-Walker, N., {et~al.} 2014, Monthly Notices of the Royal Astronomical Society, 445, 330

\bibitem[{{Hotan} {et~al.}(2021){Hotan}, {Bunton}, {Chippendale}, {Whiting}, {Tuthill}, {Moss}, {McConnell}, {Amy}, {Huynh}, {Allison}, {Anderson}, {Bannister}, {Bastholm}, {Beresford}, {Bock}, {Bolton}, {Chapman}, {Chow}, {Collier}, {Cooray}, {Cornwell}, {Diamond}, {Edwards}, {Feain}, {Franzen}, {George}, {Gupta}, {Hampson}, {Harvey-Smith}, {Hayman}, {Heywood}, {Jacka}, {Jackson}, {Jackson}, {Jeganathan}, {Johnston}, {Kesteven}, {Kleiner}, {Koribalski}, {Lee-Waddell}, {Lenc}, {Lensson}, {Mackay}, {Mahony}, {McClure-Griffiths}, {McConigley}, {Mirtschin}, {Ng}, {Norris}, {Pearce}, {Phillips}, {Pilawa}, {Raja}, {Reynolds}, {Roberts}, {Roxby}, {Sadler}, {Shields}, {Schinckel}, {Serra}, {Shaw}, {Sweetnam}, {Troup}, {Tzioumis}, {Voronkov}, \& {Westmeier}}]{askap-hotan}
{Hotan}, A.~W., {Bunton}, J.~D., {Chippendale}, A.~P., {et~al.} 2021, \pasa, 38, e009

\bibitem[{{Hudson} {et~al.}(2010){Hudson}, {Mittal}, {Reiprich}, {Nulsen}, {Andernach}, \& {Sarazin}}]{cool_core_temp_drop_2010A&A...513A..37H}
{Hudson}, D.~S., {Mittal}, R., {Reiprich}, T.~H., {et~al.} 2010, \aap, 513, A37

\bibitem[{{Hurley-Walker} {et~al.}(2021){Hurley-Walker}, {Payne}, {Filipovi{\'c}}, \& {Tothill}}]{2021map..book....2H}
{Hurley-Walker}, N., {Payne}, J.~L., {Filipovi{\'c}}, M.~D., \& {Tothill}, N. 2021, in Multimessenger Astronomy in Practice: Celestial Sources in Action, ed. M.~D. {Filipovi{\'c}} \& N.~F.~H. {Tothill} ({IOP Publishing}), 2--1

\bibitem[{{Jaffe}(1977)}]{jaffe-1977ApJ...212....1J}
{Jaffe}, W.~J. 1977, \apj, 212, 1

\bibitem[{{Johnston} {et~al.}(2007){Johnston}, {Bailes}, {Bartel}, {Baugh}, {Bietenholz}, {Blake}, {Braun}, {Brown}, {Chatterjee}, {Darling}, {Deller}, {Dodson}, {Edwards}, {Ekers}, {Ellingsen}, {Feain}, {Gaensler}, {Haverkorn}, {Hobbs}, {Hopkins}, {Jackson}, {James}, {Joncas}, {Kaspi}, {Kilborn}, {Koribalski}, {Kothes}, {Landecker}, {Lenc}, {Lovell}, {Macquart}, {Manchester}, {Matthews}, {McClure-Griffiths}, {Norris}, {Pen}, {Phillips}, {Power}, {Protheroe}, {Sadler}, {Schmidt}, {Stairs}, {Staveley-Smith}, {Stil}, {Taylor}, {Tingay}, {Tzioumis}, {Walker}, {Wall}, \& {Wolleben}}]{Science_With_ASKAP_2007PASA...24..174J}
{Johnston}, S., {Bailes}, M., {Bartel}, N., {et~al.} 2007, \pasa, 24, 174

\bibitem[{{Johnston} {et~al.}(2008){Johnston}, {Taylor}, {Bailes}, {Bartel}, {Baugh}, {Bietenholz}, {Blake}, {Braun}, {Brown}, {Chatterjee}, {Darling}, {Deller}, {Dodson}, {Edwards}, {Ekers}, {Ellingsen}, {Feain}, {Gaensler}, {Haverkorn}, {Hobbs}, {Hopkins}, {Jackson}, {James}, {Joncas}, {Kaspi}, {Kilborn}, {Koribalski}, {Kothes}, {Landecker}, {Lenc}, {Lovell}, {Macquart}, {Manchester}, {Matthews}, {McClure-Griffiths}, {Norris}, {Pen}, {Phillips}, {Power}, {Protheroe}, {Sadler}, {Schmidt}, {Stairs}, {Staveley-Smith}, {Stil}, {Tingay}, {Tzioumis}, {Walker}, {Wall}, \& {Wolleben}}]{johnston08}
{Johnston}, S., {Taylor}, R., {Bailes}, M., {et~al.} 2008, Experimental Astronomy, 22, 151

\bibitem[{{Jones} {et~al.}(2009){Jones}, {Read}, {Saunders}, {Colless}, {Jarrett}, {Parker}, {Fairall}, {Mauch}, {Sadler}, {Watson}, {Burton}, {Campbell}, {Cass}, {Croom}, {Dawe}, {Fiegert}, {Frankcombe}, {Hartley}, {Huchra}, {James}, {Kirby}, {Lahav}, {Lucey}, {Mamon}, {Moore}, {Peterson}, {Prior}, {Proust}, {Russell}, {Safouris}, {Wakamatsu}, {Westra}, \& {Williams}}]{2009MNRAS.399..683J}
{Jones}, D.~H., {Read}, M.~A., {Saunders}, W., {et~al.} 2009, \mnras, 399, 683

\bibitem[{{Jones} {et~al.}(2017){Jones}, {Nolting}, {O'Neill}, \& {Mendygral}}]{radio_channel_2017PhPl...24d1402J}
{Jones}, T.~W., {Nolting}, C., {O'Neill}, B.~J., \& {Mendygral}, P.~J. 2017, Physics of Plasmas, 24, 041402

\bibitem[{{Kalberla} {et~al.}(2005){Kalberla}, {Burton}, {Hartmann}, {Arnal}, {Bajaja}, {Morras}, \& {P{\"o}ppel}}]{LAB_survey_2005A&A...440..775K}
{Kalberla}, P.~M.~W., {Burton}, W.~B., {Hartmann}, D., {et~al.} 2005, \aap, 440, 775

\bibitem[{{Kempner} {et~al.}(2004){Kempner}, {Blanton}, {Clarke}, {En{\ss}lin}, {Johnston-Hollitt}, \& {Rudnick}}]{taxonomy_kempner_2004rcfg.proc..335K}
{Kempner}, J.~C., {Blanton}, E.~L., {Clarke}, T.~E., {et~al.} 2004, in The Riddle of Cooling Flows in Galaxies and Clusters of galaxies, ed. T.~{Reiprich}, J.~{Kempner}, \& N.~{Soker}, 335

\bibitem[{{Knowles} {et~al.}(2022){Knowles}, {Cotton}, {Rudnick}, {Camilo}, {Goedhart}, {Deane}, {Ramatsoku}, {Bietenholz}, {Br{\"u}ggen}, {Button}, {Chen}, {Chibueze}, {Clarke}, {de Gasperin}, {Ianjamasimanana}, {J{\'o}zsa}, {Hilton}, {Kesebonye}, {Kolokythas}, {Kraan-Korteweg}, {Lawrie}, {Lochner}, {Loubser}, {Marchegiani}, {Mhlahlo}, {Moodley}, {Murphy}, {Namumba}, {Oozeer}, {Parekh}, {Pillay}, {Passmoor}, {Ramaila}, {Ranchod}, {Retana-Montenegro}, {Sebokolodi}, {Sikhosana}, {Smirnov}, {Thorat}, {Venturi}, {Abbott}, {Adam}, {Adams}, {Aldera}, {Bauermeister}, {Bennett}, {Bode}, {Botha}, {Botha}, {Brederode}, {Buchner}, {Burger}, {Cheetham}, {de Villiers}, {Dikgale-Mahlakoana}, {du Toit}, {Esterhuyse}, {Fadana}, {Fanaroff}, {Fataar}, {Foley}, {Fourie}, {Frank}, {Gamatham}, {Gatsi}, {Geyer}, {Gouws}, {Gumede}, {Heywood}, {Hlakola}, {Hokwana}, {Hoosen}, {Horn}, {Horrell}, {Hugo}, {Isaacson}, {Jonas}, {Jordaan}, {Joubert}, {Julie}, {Kapp}, {Kasper}, {Kenyon}, {Kotz{\'e}}, {Kotze}, {Kriek}, {Kriel}, {Krishnan}, {Kusel}, {Legodi}, {Lehmensiek}, {Liebenberg}, {Lord}, {Lunsky}, {Madisa}, {Magnus}, {Main}, {Makhaba}, {Makhathini}, {Malan}, {Manley}, {Marais}, {Maree}, {Martens}, {Mauch}, {McAlpine}, {Merry}, {Millenaar}, {Mokone}, {Monama}, {Mphego}, {New}, {Ngcebetsha}, {Ngoasheng}, {Ockards}, {Otto}, {Patel}, {Peens-Hough}, {Perkins}, {Ramanujam}, {Ramudzuli}, {Ratcliffe}, {Renil}, {Robyntjies}, {Rust}, {Salie}, {Sambu}, {Schollar}, {Schwardt}, {Schwartz}, {Serylak}, {Siebrits}, {Sirothia}, {Slabber}, {Sofeya}, {Taljaard}, {Tasse}, {Tiplady}, {Toruvanda}, {Twum}, {van Balla}, {van der Byl}, {van der Merwe}, {van Dyk}, {Van Tonder}, {Van Wyk}, {Venter}, {Venter}, {Welz}, {Williams}, \& {Xaia}}]{2022A&A...657A..56K}
{Knowles}, K., {Cotton}, W.~D., {Rudnick}, L., {et~al.} 2022, \aap, 657, A56

\bibitem[{{Land} \& {Slosar}(2007)}]{land_2007PhRvD..76h7301L}
{Land}, K., \& {Slosar}, A. 2007, \prd, 76, 087301

\bibitem[{{Lauberts} \& {Valentijn}(1989)}]{bcg_eso-470-20_1989spce.book.....L}
{Lauberts}, A., \& {Valentijn}, E.~A. 1989, {The surface photometry catalogue of the ESO-Uppsala galaxies} (Garching: European Southern Observatory)

\bibitem[{{Lin} \& {Mohr}(2004)}]{K-BAND_PropertiesOfGalaxyClusters_2004ApJ...617..879L}
{Lin}, Y.-T., \& {Mohr}, J.~J. 2004, \apj, 617, 879

\bibitem[{Mauch {et~al.}(2003)Mauch, Murphy, Buttery, Curran, Hunstead, Piestrzynski, Robertson, \& Sadler}]{sumss-mauch-10.1046/j.1365-8711.2003.06605.x}
Mauch, T., Murphy, T., Buttery, H.~J., {et~al.} 2003, Monthly Notices of the Royal Astronomical Society, 342, 1117

\bibitem[{{McConnell} {et~al.}(2016){McConnell}, {Allison}, {Bannister}, {Bell}, {Bignall}, {Chippendale}, {Edwards}, {Harvey-Smith}, {Hegarty}, {Heywood}, {Hotan}, {Indermuehle}, {Lenc}, {Marvil}, {Popping}, {Raja}, {Reynolds}, {Sault}, {Serra}, {Voronkov}, {Whiting}, {Amy}, {Axtens}, {Ball}, {Bateman}, {Bock}, {Bolton}, {Brodrick}, {Brothers}, {Brown}, {Bunton}, {Cheng}, {Cornwell}, {DeBoer}, {Feain}, {Gough}, {Gupta}, {Guzman}, {Hampson}, {Hay}, {Hayman}, {Hoyle}, {Humphreys}, {Jacka}, {Jackson}, {Jackson}, {Jeganathan}, {Joseph}, {Koribalski}, {Leach}, {Lensson}, {MacLeod}, {Mackay}, {Marquarding}, {McClure-Griffiths}, {Mirtschin}, {Mitchell}, {Neuhold}, {Ng}, {Norris}, {Pearce}, {Qiao}, {Schinckel}, {Shields}, {Shimwell}, {Storey}, {Troup}, {Turner}, {Tuthill}, {Tzioumis}, {Wark}, {Westmeier}, {Wilson}, \& {Wilson}}]{mcconnell16}
{McConnell}, D., {Allison}, J.~R., {Bannister}, K., {et~al.} 2016, \pasa, 33, e042

\bibitem[{{McConnell} {et~al.}(2020){McConnell}, {Hale}, {Lenc}, {Banfield}, {Heald}, {Hotan}, {Leung}, {Moss}, {Murphy}, {O'Brien}, {Pritchard}, {Raja}, {Sadler}, {Stewart}, {Thomson}, {Whiting}, {Allison}, {Amy}, {Anderson}, {Ball}, {Bannister}, {Bell}, {Bock}, {Bolton}, {Bunton}, {Chippendale}, {Collier}, {Cooray}, {Cornwell}, {Diamond}, {Edwards}, {Gupta}, {Hayman}, {Heywood}, {Jackson}, {Koribalski}, {Lee-Waddell}, {McClure-Griffiths}, {Ng}, {Norris}, {Phillips}, {Reynolds}, {Roxby}, {Schinckel}, {Shields}, {Tremblay}, {Tzioumis}, {Voronkov}, \& {Westmeier}}]{RACS-2020PASA...37...48M}
{McConnell}, D., {Hale}, C.~L., {Lenc}, E., {et~al.} 2020, \pasa, 37, e048

\bibitem[{{Migkas} {et~al.}(2020){Migkas}, {Schellenberger}, {Reiprich}, {Pacaud}, {Ramos-Ceja}, \& {Lovisari}}]{migkas_xray_2020A&A...636A..15M}
{Migkas}, K., {Schellenberger}, G., {Reiprich}, T.~H., {et~al.} 2020, \aap, 636, A15

\bibitem[{{Murgia} {et~al.}(2010){Murgia}, {Eckert}, {Govoni}, {Ferrari}, {Pand ey-Pommier}, {Nevalainen}, \& {Paltani}}]{murgia_2010A&A...514A..76M}
{Murgia}, M., {Eckert}, D., {Govoni}, F., {et~al.} 2010, \aap, 514, A76

\bibitem[{{Nelson}(1992)}]{atca-nelson}
{Nelson}, G.~J. 1992, Journal of Electrical and Electronics Engineering Australia, 12, 113

\bibitem[{{Norris} {et~al.}(2011){Norris}, {Hopkins}, {Afonso}, {Brown}, {Condon}, {Dunne}, {Feain}, {Hollow}, {Jarvis}, {Johnston-Hollitt}, {Lenc}, {Middelberg}, {Padovani}, {Prandoni}, {Rudnick}, {Seymour}, {Umana}, {Andernach}, {Alexander}, {Appleton}, {Bacon}, {Banfield}, {Becker}, {Brown}, {Ciliegi}, {Jackson}, {Eales}, {Edge}, {Gaensler}, {Giovannini}, {Hales}, {Hancock}, {Huynh}, {Ibar}, {Ivison}, {Kennicutt}, {Kimball}, {Koekemoer}, {Koribalski}, {L{\'o}pez-S{\'a}nchez}, {Mao}, {Murphy}, {Messias}, {Pimbblet}, {Raccanelli}, {Randall}, {Reiprich}, {Roseboom}, {R{\"o}ttgering}, {Saikia}, {Sharp}, {Slee}, {Smail}, {Thompson}, {Urquhart}, {Wall}, \& {Zhao}}]{EMU_2011PASA...28..215N}
{Norris}, R.~P., {Hopkins}, A.~M., {Afonso}, J., {et~al.} 2011, \pasa, 28, 215

\bibitem[{{Norris} {et~al.}(2021){Norris}, {Marvil}, {Collier}, {Kapi{\'n}ska}, {O'Brien}, {Rudnick}, {Andernach}, {Asorey}, {Brown}, {Br{\"u}ggen}, {Crawford}, {English}, {Rahman}, {Filipovi{\'c}}, {Gordon}, {G{\"u}rkan}, {Hale}, {Hopkins}, {Huynh}, {HyeongHan}, {James Jee}, {Koribalski}, {Lenc}, {Luken}, {Parkinson}, {Prandoni}, {Raja}, {Reiprich}, {Riseley}, {Shabala}, {Sheil}, {Vernstrom}, {Whiting}, {Allison}, {Anderson}, {Ball}, {Bell}, {Bunton}, {Galvin}, {Gupta}, {Hotan}, {Jacka}, {Macgregor}, {Mahony}, {Maio}, {Moss}, {Pandey-Pommier}, \& {Voronkov}}]{norris20}
{Norris}, R.~P., {Marvil}, J., {Collier}, J.~D., {et~al.} 2021, \pasa, 38, e046

\bibitem[{{Offringa} {et~al.}(2016){Offringa}, {Trott}, {Hurley-Walker}, {Johnston-Hollitt}, {McKinley}, {Barry}, {Beardsley}, {Bowman}, {Briggs}, {Carroll}, {Dillon}, {Ewall-Wice}, {Feng}, {Gaensler}, {Greenhill}, {Hazelton}, {Hewitt}, {Jacobs}, {Kim}, {Kittiwisit}, {Lenc}, {Line}, {Loeb}, {Mitchell}, {Morales}, {Neben}, {Paul}, {Pindor}, {Pober}, {Procopio}, {Riding}, {Sethi}, {Shankar}, {Subrahmanyan}, {Sullivan}, {Tegmark}, {Thyagarajan}, {Tingay}, {Wayth}, {Webster}, \& {Wyithe}}]{oth+16}
{Offringa}, A.~R., {Trott}, C.~M., {Hurley-Walker}, N., {et~al.} 2016, \mnras, 458, 1057

\bibitem[{Olowin(1988)}]{Abell_Southern_Extension_1988_PASP_100_1354_10.2307/40679228}
Olowin, R.~P. 1988, Publications of the Astronomical Society of the Pacific, 100, 1354

\bibitem[{{Onken} {et~al.}(2019){Onken}, {Wolf}, {Bessell}, {Chang}, {Da Costa}, {Luvaul}, {Mackey}, {Schmidt}, \& {Shao}}]{skymapper_dr2_2019PASA...36...33O}
{Onken}, C.~A., {Wolf}, C., {Bessell}, M.~S., {et~al.} 2019, \pasa, 36, e033

\bibitem[{Ozawa {et~al.}(2015)Ozawa, Nakanishi, Akahori, Anraku, Takizawa, Takahashi, Onodera, Tsuda, \& Sofue}]{Ozawa_2015}
Ozawa, T., Nakanishi, H., Akahori, T., {et~al.} 2015, Publications of the Astronomical Society of Japan, 67, 110

\bibitem[{{Pasini} {et~al.}(2022){Pasini}, {Edler}, {Br{\"u}ggen}, {de Gasperin}, {Botteon}, {Rajpurohit}, {van Weeren}, {Gastaldello}, {Gaspari}, {Brunetti}, {Cuciti}, {Nanci}, {di Gennaro}, {Rossetti}, {Dallacasa}, {Hoang}, \& {Riseley}}]{Pasini2022_A1550}
{Pasini}, T., {Edler}, H.~W., {Br{\"u}ggen}, M., {et~al.} 2022, \aap, 663, A105

\bibitem[{{Paul} {et~al.}(2023){Paul}, {Kale}, {Datta}, {Basu}, {Sur}, {Parekh}, {Gupta}, {Chatterjee}, {Salunkhe}, {Iqbal}, {Pandey-Pommier}, {Raja}, {Rahaman}, {Raychaudhury}, {Nath}, \& {Majumdar}}]{Paul_2023JApA...44...38P}
{Paul}, S., {Kale}, R., {Datta}, A., {et~al.} 2023, Journal of Astrophysics and Astronomy, 44, 38

\bibitem[{{Piffaretti} {et~al.}(2011){Piffaretti}, {Arnaud}, {Pratt}, {Pointecouteau}, \& {Melin}}]{mcxc-2011A&A...534A.109P}
{Piffaretti}, R., {Arnaud}, M., {Pratt}, G.~W., {Pointecouteau}, E., \& {Melin}, J.~B. 2011, \aap, 534, A109

\bibitem[{{Pizzo}(2011)}]{2011JApA...32..567P}
{Pizzo}, R.~F. 2011, Journal of Astrophysics and Astronomy, 32, 567

\bibitem[{{Rajpurohit} {et~al.}(2022){Rajpurohit}, {Hoeft}, {Wittor}, {van Weeren}, {Vazza}, {Rudnick}, {Rajpurohit}, {Forman}, {Riseley}, {Brienza}, {Bonafede}, {Rajpurohit}, {Dom{\'\i}nguez-Fern{\'a}ndez}, {Eilek}, {Bonnassieux}, {Br{\"u}ggen}, {Loi}, {R{\"o}ttgering}, {Drabent}, {Locatelli}, {Botteon}, {Brunetti}, \& {Clarke}}]{Rajpurohit2021-macj0717}
{Rajpurohit}, K., {Hoeft}, M., {Wittor}, D., {et~al.} 2022, \aap, 657, A2

\bibitem[{{Ramos-Ceja} {et~al.}(2019){Ramos-Ceja}, {Pacaud}, {Reiprich}, {Migkas}, {Lovisari}, \& {Schellenberger}}]{2019A&A...626A..48R}
{Ramos-Ceja}, M.~E., {Pacaud}, F., {Reiprich}, T.~H., {et~al.} 2019, \aap, 626, A48

\bibitem[{{Read} {et~al.}(2001){Read}, {Filipovi{\'c}}, {Pietsch}, \& {Jones}}]{S0102_2001A&A...369..467R}
{Read}, A.~M., {Filipovi{\'c}}, M.~D., {Pietsch}, W., \& {Jones}, P.~A. 2001, \aap, 369, 467

\bibitem[{{Reiprich}(2017)}]{eeHIFLUGCS_2017xru..conf..189R}
{Reiprich}, T. 2017, in The X-ray Universe 2017, ed. J.-U. {Ness} \& S.~{Migliari}, 189

\bibitem[{{Riseley} {et~al.}(2022{\natexlab{a}}){Riseley}, {Rajpurohit}, {Loi}, {Botteon}, {Timmerman}, {Biava}, {Bonafede}, {Bonnassieux}, {Brunetti}, {En{\ss}lin}, {Di Gennaro}, {Ignesti}, {Shimwell}, {Stuardi}, {Vernstrom}, \& {van Weeren}}]{Riseley2022-ms1455}
{Riseley}, C.~J., {Rajpurohit}, K., {Loi}, F., {et~al.} 2022{\natexlab{a}}, \mnras, 512, 4210

\bibitem[{{Riseley} {et~al.}(2022{\natexlab{b}}){Riseley}, {Bonnassieux}, {Vernstrom}, {Galvin}, {Chokshi}, {Botteon}, {Rajpurohit}, {Duchesne}, {Bonafede}, {Rudnick}, {Hoeft}, {Quici}, {Eckert}, {Brienza}, {Tasse}, {Carretti}, {Collier}, {Diego}, {Di Mascolo}, {Hopkins}, {Johnston-Hollitt}, {Keel}, {Koribalski}, \& {Reiprich}}]{Riseley2022_Abell3266}
{Riseley}, C.~J., {Bonnassieux}, E., {Vernstrom}, T., {et~al.} 2022{\natexlab{b}}, \mnras, 515, 1871

\bibitem[{{Riseley} {et~al.}(2023){Riseley}, {Biava}, {Lusetti}, {Bonafede}, {Bonnassieux}, {Botteon}, {Loi}, {Brunetti}, {Cassano}, {Osinga}, {Rajpurohit}, {R{\"o}ttgering}, {Shimwell}, {Timmerman}, \& {van Weeren}}]{Riseley2023-a1413}
{Riseley}, C.~J., {Biava}, N., {Lusetti}, G., {et~al.} 2023, \mnras, 524, 6052

\bibitem[{{Riseley} {et~al.}(2024){Riseley}, {Bonafede}, {Bruno}, {Botteon}, {Rossetti}, {Biava}, {Bonnassieux}, {Loi}, {Vernstrom}, \& {Balboni}}]{Riseley2024-a2142}
{Riseley}, C.~J., {Bonafede}, A., {Bruno}, L., {et~al.} 2024, arXiv e-prints, arXiv:2403.00414

\bibitem[{{Rudnick} {et~al.}(2022{\natexlab{a}}){Rudnick}, {Br{\"u}ggen}, {Brunetti}, {Cotton}, {Forman}, {Jones}, {Nolting}, {Schellenberger}, \& {van Weeren}}]{2022ApJ...935..168R}
{Rudnick}, L., {Br{\"u}ggen}, M., {Brunetti}, G., {et~al.} 2022{\natexlab{a}}, \apj, 935, 168

\bibitem[{{Rudnick} {et~al.}(2022{\natexlab{b}}){Rudnick}, {Br{\"u}ggen}, {Brunetti}, {Cotton}, {Forman}, {Jones}, {Nolting}, {Schellenberger}, \& {van Weeren}}]{A194}
---. 2022{\natexlab{b}}, \apj, 935, 168

\bibitem[{{Sault} {et~al.}(1995){Sault}, {Teuben}, \& {Wright}}]{miriad_1995ASPC...77..433S}
{Sault}, R.~J., {Teuben}, P.~J., \& {Wright}, M.~C.~H. 1995, in Astronomical Society of the Pacific Conference Series, Vol.~77, Astronomical Data Analysis Software and Systems IV, ed. R.~A. {Shaw}, H.~E. {Payne}, \& J.~J.~E. {Hayes}, 433

\bibitem[{{Savini} {et~al.}(2018){Savini}, {Bonafede}, {Br{\"u}ggen}, {van Weeren}, {Brunetti}, {Intema}, {Botteon}, {Shimwell}, {Wilber}, {Rafferty}, {Giacintucci}, {Cassano}, {Cuciti}, {de Gasperin}, {R{\"o}ttgering}, {Hoeft}, \& {White}}]{savini-2018MNRAS.478.2234S}
{Savini}, F., {Bonafede}, A., {Br{\"u}ggen}, M., {et~al.} 2018, \mnras, 478, 2234

\bibitem[{{Savini} {et~al.}(2019){Savini}, {Bonafede}, {Br{\"u}ggen}, {Rafferty}, {Shimwell}, {Botteon}, {Brunetti}, {Intema}, {Wilber}, {Cassano}, {Vazza}, {van Weeren}, {Cuciti}, {De Gasperin}, {R{\"o}ttgering}, {Sommer}, {B{\^\i}rzan}, \& {Drabent}}]{2019A&A...622A..24S}
---. 2019, \aap, 622, A24

\bibitem[{{Slee} {et~al.}(2001){Slee}, {Roy}, {Murgia}, {Andernach}, \& {Ehle}}]{Slee_AngResClusters_2001AJ....122.1172S}
{Slee}, O.~B., {Roy}, A.~L., {Murgia}, M., {Andernach}, H., \& {Ehle}, M. 2001, \aj, 122, 1172

\bibitem[{{Stroe} {et~al.}(2014){Stroe}, {Sobral}, {R{\"o}ttgering}, \& {van Weeren}}]{Stroe-10.1093/mnras/stt2286}
{Stroe}, A., {Sobral}, D., {R{\"o}ttgering}, H. J.~A., \& {van Weeren}, R.~J. 2014, \mnras, 438, 1377

\bibitem[{{Thompson} {et~al.}(2017){Thompson}, {Moran}, \& {Swenson}}]{2017isra.book.....T}
{Thompson}, A.~R., {Moran}, J.~M., \& {Swenson}, George~W., J. 2017, {Interferometry and Synthesis in Radio Astronomy, 3rd Edition} (Springer International Publishing AG), doi:\url{10.1007/978-3-319-44431-4}

\bibitem[{{Tingay} {et~al.}(2013){Tingay}, {Goeke}, {Bowman}, {Emrich}, {Ord}, {Mitchell}, {Morales}, {Booler}, {Crosse}, {Wayth}, {Lonsdale}, {Tremblay}, {Pallot}, {Colegate}, {Wicenec}, {Kudryavtseva}, {Arcus}, {Barnes}, {Bernardi}, {Briggs}, {Burns}, {Bunton}, {Cappallo}, {Corey}, {Deshpande}, {Desouza}, {Gaensler}, {Greenhill}, {Hall}, {Hazelton}, {Herne}, {Hewitt}, {Johnston-Hollitt}, {Kaplan}, {Kasper}, {Kincaid}, {Koenig}, {Kratzenberg}, {Lynch}, {Mckinley}, {Mcwhirter}, {Morgan}, {Oberoi}, {Pathikulangara}, {Prabu}, {Remillard}, {Rogers}, {Roshi}, {Salah}, {Sault}, {Udaya-Shankar}, {Schlagenhaufer}, {Srivani}, {Stevens}, {Subrahmanyan}, {Waterson}, {Webster}, {Whitney}, {Williams}, {Williams}, \& {Wyithe}}]{2013PASA...30....7T}
{Tingay}, S.~J., {Goeke}, R., {Bowman}, J.~D., {et~al.} 2013, \pasa, 30, e007

\bibitem[{{T{\"u}mer} {et~al.}(2023){T{\"u}mer}, {Wik}, {Zhang}, {Hoang}, {Gaspari}, {van Weeren}, {Rudnick}, {Stuardi}, {Mernier}, {Simionescu}, {Rojas Bolivar}, {Kraft}, {Akamatsu}, \& {de Plaa}}]{tumer_halos_and_mergers_2023ApJ...942...79T}
{T{\"u}mer}, A., {Wik}, D.~R., {Zhang}, X., {et~al.} 2023, \apj, 942, 79

\bibitem[{{van Weeren} {et~al.}(2019){van Weeren}, {de Gasperin}, {Akamatsu}, {Br{\"u}ggen}, {Feretti}, {Kang}, {Stroe}, \& {Zandanel}}]{van-weeren-2019SSRv..215...16V}
{van Weeren}, R.~J., {de Gasperin}, F., {Akamatsu}, H., {et~al.} 2019, \ssr, 215, 16

\bibitem[{{van Weeren} {et~al.}(2011){van Weeren}, {Hoeft}, {R{\"o}ttgering}, {Br{\"u}ggen}, {Intema}, \& {van Velzen}}]{van-weeren-2011A&A...528A..38V}
{van Weeren}, R.~J., {Hoeft}, M., {R{\"o}ttgering}, H.~J.~A., {et~al.} 2011, \aap, 528, A38

\bibitem[{{van Weeren} {et~al.}(2010){van Weeren}, {R{\"o}ttgering}, {Br{\"u}ggen}, \& {Hoeft}}]{2010Sci...330..347V}
{van Weeren}, R.~J., {R{\"o}ttgering}, H. J.~A., {Br{\"u}ggen}, M., \& {Hoeft}, M. 2010, Science, 330, 347

\bibitem[{{van Weeren} {et~al.}(2016){van Weeren}, {Brunetti}, {Br{\"u}ggen}, {Andrade-Santos}, {Ogrean}, {Williams}, {R{\"o}ttgering}, {Dawson}, {Forman}, {de Gasperin}, {Hardcastle}, {Jones}, {Miley}, {Rafferty}, {Rudnick}, {Sabater}, {Sarazin}, {Shimwell}, {Bonafede}, {Best}, {B{\^\i}rzan}, {Cassano}, {Chy{\.z}y}, {Croston}, {Dijkema}, {En{\ss}lin}, {Ferrari}, {Heald}, {Hoeft}, {Horellou}, {Jarvis}, {Kraft}, {Mevius}, {Intema}, {Murray}, {Orr{\'u}}, {Pizzo}, {Sridhar}, {Simionescu}, {Stroe}, {van der Tol}, \& {White}}]{toothbrush-2016ApJ...818..204V}
{van Weeren}, R.~J., {Brunetti}, G., {Br{\"u}ggen}, M., {et~al.} 2016, \apj, 818, 204

\bibitem[{{Velovi{\'c}} {et~al.}(2023){Velovi{\'c}}, {Cotton}, {Filipovi{\'c}}, {Norris}, {Barnes}, \& {Condon}}]{velovic_2023MNRAS.523.1933V}
{Velovi{\'c}}, V., {Cotton}, W.~D., {Filipovi{\'c}}, M.~D., {et~al.} 2023, \mnras, 523, 1933

\bibitem[{{Venturi} {et~al.}(2017){Venturi}, {Bardelli}, {Dallacasa}, {Di Gennaro}, {Gastaldello}, {Giacintucci}, \& {Rossetti}}]{venturi_shapleyCC_2017Galax...5...16V}
{Venturi}, T., {Bardelli}, S., {Dallacasa}, D., {et~al.} 2017, Galaxies, 5, 16

\bibitem[{{Wayth} {et~al.}(2018){Wayth}, {Tingay}, {Trott}, {Emrich}, {Johnston-Hollitt}, {McKinley}, {Gaensler}, {Beardsley}, {Booler}, {Crosse}, {Franzen}, {Horsley}, {Kaplan}, {Kenney}, {Morales}, {Pallot}, {Sleap}, {Steele}, {Walker}, {Williams}, {Wu}, {Cairns}, {Filipovic}, {Johnston}, {Murphy}, {Quinn}, {Staveley-Smith}, {Webster}, \& {Wyithe}}]{wtt+18_v2_2018PASA...35...33W}
{Wayth}, R.~B., {Tingay}, S.~J., {Trott}, C.~M., {et~al.} 2018, \pasa, 35, e033

\bibitem[{{Wilson} {et~al.}(2011){Wilson}, {Ferris}, {Axtens}, {Brown}, {Davis}, {Hampson}, {Leach}, {Roberts}, {Saunders}, {Koribalski}, {Caswell}, {Lenc}, {Stevens}, {Voronkov}, {Wieringa}, {Brooks}, {Edwards}, {Ekers}, {Emonts}, {Hindson}, {Johnston}, {Maddison}, {Mahony}, {Malu}, {Massardi}, {Mao}, {McConnell}, {Norris}, {Schnitzeler}, {Subrahmanyan}, {Urquhart}, {Thompson}, \& {Wark}}]{wilson_cabb_2011MNRAS.416..832W}
{Wilson}, W.~E., {Ferris}, R.~H., {Axtens}, P., {et~al.} 2011, \mnras, 416, 832

\bibitem[{{Wright} {et~al.}(2010){Wright}, {Eisenhardt}, {Mainzer}, {Ressler}, {Cutri}, {Jarrett}, {Kirkpatrick}, {Padgett}, {McMillan}, {Skrutskie}, {Stanford}, {Cohen}, {Walker}, {Mather}, {Leisawitz}, {Gautier}, {McLean}, {Benford}, {Lonsdale}, {Blain}, {Mendez}, {Irace}, {Duval}, {Liu}, {Royer}, {Heinrichsen}, {Howard}, {Shannon}, {Kendall}, {Walsh}, {Larsen}, {Cardon}, {Schick}, {Schwalm}, {Abid}, {Fabinsky}, {Naes}, \& {Tsai}}]{wise_2010AJ....140.1868W}
{Wright}, E.~L., {Eisenhardt}, P. R.~M., {Mainzer}, A.~K., {et~al.} 2010, \aj, 140, 1868

\end{thebibliography}
%%%%%%%%%
\appendix
%%%%%%%%%
\end{document}